\documentclass[journal,10pt,draftclsnofoot,onecolumn]{IEEEtran}
\IEEEoverridecommandlockouts

\usepackage{cite}
\usepackage{amsmath,amssymb,amsfonts,mathrsfs}
\usepackage{bbm}

\usepackage[normalem]{ulem}
\newtheorem{theorem}{Theorem}
\newtheorem{lemma}{Lemma}
\newtheorem{corollary}{Corollary}

\newtheorem{proposition}{Proposition}
\newtheorem{definition}{Definition}

\usepackage{placeins}
\usepackage{graphicx}
\usepackage{textcomp}
\usepackage{xcolor}
\usepackage[bookmarks=false]{hyperref}
\usepackage[nameinlink,noabbrev]{cleveref}

\crefname{theorem}{theorem}{theorems}
\Crefname{theorem}{Theorem}{Theorems}
\crefname{lemma}{lemma}{lemmas}
\Crefname{lemma}{Lemma}{Lemmas}
\crefname{corollary}{corollary}{corollaries}
\Crefname{corollary}{Corollary}{Corollaries}
\crefname{proposition}{proposition}{propositions}
\Crefname{proposition}{Proposition}{Propositions}
\crefname{definition}{definition}{definitions}
\Crefname{definition}{Definition}{Definitions}
\crefname{remark}{remark}{remarks}
\Crefname{remark}{Remark}{Remarks}
\crefname{assumption}{assumption}{assumptions}
\Crefname{assumption}{Assumption}{Assumptions}
\crefname{problem}{problem}{problems}
\Crefname{problem}{Problem}{Problems}
\crefname{section}{section}{sections}
\Crefname{section}{Section}{Sections}
\crefname{subsection}{section}{sections}
\Crefname{subsection}{Section}{Sections}
\crefname{equation}{equation}{equations}
\Crefname{equation}{Equation}{Equations}
\crefname{figure}{Fig.}{Figs.}
\Crefname{figure}{Fig.}{Figs.}
\crefname{appendix}{appendix}{appendices}
\Crefname{appendix}{Appendix}{Appendices}

\def\BibTeX{{\rm B\kern-.05em{\sc i\kern-.025em b}\kern-.08em
    T\kern-.1667em\lower.7ex\hbox{E}\kern-.125emX}}

\begin{document}

\title{Finite Sample Bounds for Composite Hypothesis Testing}
\author{\IEEEauthorblockN{El{\'i}as Vera-Sig{\"u}enza, \and Amedeo Roberto Esposito}\\
\IEEEauthorblockA{
Okinawa Institute of Science and Technology (OIST) \\
Onna, Okinawa, Japan\\
 \{elias.vera, amedeo esposito\}@oist.jp} 
 }

\maketitle

\begin{abstract}
We investigate composite binary hypothesis testing in the finite sample regime under asymmetric error constraints. Using R\'enyi divergences, we derive explicit
achievability and converse bounds for the optimal Type II error. When the Type I error is constrained to decay exponentially with sample size, the bounds identify a phase transition and yield a strong converse above it. In the composite problem, the phase transition threshold is given by the joint KL projection over the alternative and null classes. Achievability is obtained through a joint R\'enyi projection whose log likelihood ratio defines a single test with uniform error control over both hypothesis classes, without requiring the projected pair to be least favourable. For compact convex classes with full support on a finite alphabet, we determine the exact error exponents on both sides of the transition and show that the achievable exponent is attained at a unique R\'enyi order. The same framework recovers the fixed Type I composite Chernoff--Stein exponent and yields a polynomial refinement of the finite sample achievability result. We further identify conditions under which the projected pair is least favourable at finite sample size.
\end{abstract}
\begin{IEEEkeywords}
Composite Hypothesis Testing, Minimax Testing, Finite Sample, R\'enyi Divergence, R\'enyi projection, Type II Error Exponent, Strong Converse, Least Favourable Distributions, Robust Detection
\end{IEEEkeywords}

%##################################################
\section{Introduction}
\label{sec:introduction}

Binary hypothesis testing asks whether observed data were generated under a null hypothesis or an alternative hypothesis. In the Neyman Pearson (NP) formulation, the probability of rejecting a true null hypothesis is constrained, while the probability of accepting a false null hypothesis is minimised \cite{neyman1933testing,lehmann2005testing}. These are the Type I and Type II errors, respectively.

When the null and alternative hypotheses each specify a single probability law, the problem is a simple hypothesis test, and the NP lemma identifies an optimal likelihood ratio test~\cite{neyman1933testing}. Under a fixed Type I error constraint, the asymptotic behaviour of the optimal Type II error is characterised by the Chernoff-Stein lemma, which gives the Kullback Leibler (KL) divergence as the optimal Type II error exponent~\cite{cover2006elements}. If instead the Type I error is required to decay exponentially as a function of sample size, the asymptotic behaviour depends on an imposed rate $r$. Here the problem exhibits a phase transition. The Type II error decays exponentially to zero when $r$ is below the phase transition threshold, while above it the Type II error converges exponentially to one. The Type II error decays exponentially to zero below a critical Type I error rate, while above this rate it converges exponentially to one. The threshold separating these two regimes is the KL divergence from the alternative hypothesis to the null hypothesis~\cite{blahut1974hypothesis,han2002strong}. Bruno et al.~\cite{bruno2026finite} recently derived finite sample achievability and converse bounds for the simple binary hypothesis testing problem using R\'enyi divergences and recovered this classical threshold rate. They thus provided an explicit achievable exponential rate for the Type II error below it.

In this work, we study the problem when both hypotheses are composite. The null and alternative hypotheses each comprise a class of probability laws, and the test must be chosen without knowing which member of either class generated the data. Consequently, the Type I error constraint must hold for every distribution in the null class, while the Type II error is assessed over the alternative class. Unlike in simple hypothesis testing, the hypotheses no longer determine a single likelihood ratio because they do not specify a single null and alternative distribution. One standard approach is the generalised likelihood ratio test, which compares the maximised likelihoods under the two classes~\cite{zeitouni1992generalized}.Although widely used, the GLR test is not generally optimal for composite hypothesis testing~\cite{zeitouni1992generalized}. Another classical approach uses least favourable distributions to reduce a composite testing problem to a simple binary test~\cite{fauss2021minimax}.

Robust detection provides another classical approach. Here one seeks least favourable distributions that reduce the composite problem to a simple binary test~\cite{fauss2021minimax}. Under structural conditions such as stochastic ordering, Choquet capacities, or Blackwell dominance, a pair can be identified that simultaneously attains the extremal Type I and Type II errors over the two classes \cite{huber1965robust,verdu2003minimax,fauss2021minimax}. The resulting likelihood ratio test is then NP optimal for the composite problem at every sample size. Such a pair is not guaranteed to exist for general composite classes, and classical existence results require additional structural conditions. Finite sample least favourability is also stronger than determining the optimal asymptotic error exponent. A pair may determine this exponent without attaining the extremal error probabilities at finite sample size. G\"ul~\cite{gul2026structural} illustrates this distinction for convex uncertainty classes.

Other studies characterise asymptotic error exponents for structured composite problems. Tomamichel $\&$ Hayashi \cite{tomamichel2017operational} consider asymmetric testing with a fixed i.i.d. null distribution and a structured composite alternative. Their framework uses a sequence of universal distributions on the sample space that dominates the admissible alternatives up to a factor that grows polynomially with sample size. This construction permits a single test to control the composite alternative and yields the critical exponential Type II error rate separating the error exponent and strong converse regimes, together with the corresponding exact exponents.

For product and Markov alternative classes, these exponents admit operational interpretations in terms of variants of Sibson-$\alpha$ mutual information and R\'enyi conditional mutual information. They also derive a Gaussian second order expansion around this critical rate on the $\sqrt n$ scale. This concerns local deviations around the critical rate and is distinct from the fixed large deviation refinement considered here, where polynomial factors modify the dominant exponential behaviour.

Shayevitz~\cite{shayevitz2011renyi} gives a different operational interpretation of R\'enyi divergence through a two sensor composite testing problem, showing under an extremal noise condition that the optimal miss detection exponent is determined by a R\'enyi divergence between the corresponding distribution families. Huang and Moulin~\cite{huang2014strong} study a simple null against a finite collection of product alternatives and derive strong large deviation asymptotics for the achievable error region, determining it up to constant order. Moulin and Johnstone~\cite{moulin2015strong} consider a known null against a regular finite dimensional exponential family alternative and obtain sharp asymptotics for Rao and generalised likelihood ratio tests, including dimension dependent higher order terms. In the quantum setting, Berta, Brand\~ao, and Hirche~\cite{berta2021composite} extend Stein's lemma to convex combinations of i.i.d. quantum states and show that the optimal Type II error exponent is generally given by a regularised quantum relative entropy, which reduces to a nonregularised expression in particular cases. Mosonyi, Szil\'agyi, and Weiner~\cite{mosonyi2021error} give a broader analysis of composite error exponents, identifying conditions under which the optimal composite exponent coincides with the worst pairwise exponent and examples in which this reduction fails.

These results address important structured instances of composite testing, but none provide finite sample achievability and converse bounds for the classical setting considered here, where both hypotheses are composite and a single test is subject to a uniform Type I error constraint over the null class while the Type II error is evaluated over the alternative class. Our work provides explicit bounds for this problem.

%###################################
\subsection{Contributions}
Let $\mathcal C_0$ and $\mathcal C_1$ denote the null and alternative classes, and write $\beta_n^\star(\varepsilon)$ for the optimal Type II error under the uniform Type I constraint $\varepsilon$.

Our main contributions are as follows.
\begin{enumerate}

    \item We derive a finite sample converse for the composite problem. For every admissible test and every
    $P\in\mathcal C_0$, $Q\in\mathcal C_1$, pairwise reduction gives a lower bound on $\beta_n^\star(\varepsilon;\mathcal C_0,\mathcal C_1)$
    in terms of the corresponding simple binary problem $(P,Q)$. This yields an explicit composite R\'enyi converse without assuming the existence of a least favourable pair. See \Cref{sec:composite-converse}.

    \item We derive a finite sample achievability bound without assuming that the composite problem admits a least favourable pair. For $\lambda\in(0,1)$, a joint R\'enyi projection $(Q_\lambda^\star,P_\lambda^\star)$ attaining $\inf\limits_{Q\in\mathcal C_1} \inf\limits_{P\in\mathcal C_0}D_\lambda(Q\|P)$ defines a single test through the log-likelihood ratio $\log(Q_\lambda^\star/P_\lambda^\star)$. We show that this test controls the Type I error uniformly over $\mathcal C_0$ and the Type II error uniformly over $\mathcal C_1$. Again, this does not assume least favourability of the distribution and the likelihood-ratio test matches the least favourable one only in certain cases. See \Cref{sec:composite-achievability}.  

    \item Under the exponentially decaying Type I constraint $\varepsilon=e^{-nr}$, we extend the classical phase transition for simple binary testing to the composite setting. The threshold is given by the KL projection of the alternative class onto the null class, $r_c=\inf_{Q\in\mathcal C_1}\inf_{P\in\mathcal C_0}D(Q\|P)$. For $r<r_c$, $\beta_n^\star(e^{-nr})$ decays exponentially to zero, whereas for $r>r_c$ it converges to one. For compact convex classes with full support on a finite alphabet, we determine the exact exponents on both sides of this transition. In the achievable regime, the Type II error exponent is the smallest exponent among all simple pairs. In the converse regime, the exponent governing convergence to one is the largest strong converse exponent among all simple pairs. See \Cref{subsec:threshold-rate,sec:exact-exponent,sec:exact-strong-converse-exponent}.

    \item For compact convex classes with full support on a finite alphabet, we show that our finite sample R\'enyi achievability bound recovers the known composite Chernoff--Stein exponent under a fixed Type I constraint. Specifically, the exponent $\inf_{\substack{P\in\mathcal C_0\\Q\in\mathcal C_1}}D(P\|Q)$ emerges from the order zero limit of the R\'enyi achievability expression, while the order one limit gives the phase transition threshold $\inf_{\substack{Q\in\mathcal C_1\\P\in\mathcal C_0}}D(Q\|P)$. Thus the R\'enyi formulation connects the fixed and exponentially decaying Type I regimes through its order zero and order one limits. Moreover, the achievable exponent for $\varepsilon=e^{-nr}$ converges to the fixed Type I exponent as $r\downarrow0$, giving a continuous connection between the two regimes at zero rate. See \Cref{sec:fixed-type-one}.
     
    \item We analyse the achievability result beyond the exponential rate. We show that the optimal Type II error has a polynomial prefactor $n^{-1/(2\lambda_r^\star)}$ multiplying its dominant exponential decay. We also optimise the likelihood ratio test induced by the R\'enyi projection under the exact Type I constraint, prove uniqueness of the maximising R\'enyi order $\lambda_r^\star$, and identify the corresponding logarithmic threshold correction $-(2\lambda_r^\star)^{-1}\log n+O(1)$. See \Cref{sec:exact-exponent,subsec:projected-tightening,subsec:polynomial-refinement}.

    \item We ask whether the pair $(Q_\lambda^\star,P_\lambda^\star)$, selected by the R\'enyi projection, can also be least favourable at finite sample size. We give sufficient stochastic ordering conditions on $\log(Q_\lambda^\star/P_\lambda^\star)$ under which the composite problem reduces exactly to the corresponding simple binary problem. We then show that these conditions hold for separated one parameter natural exponential families, where the R\'enyi projection gives the least favourable pair. See \Cref{sec:exact-reduction}.

\end{enumerate}

%##################################################
\section{Problem formulation and definitions}
\label{sec:problem-formulation}

%##################################################
\subsection{Problem formulation}
Let $\mathcal C_0$ and $\mathcal C_1$ be nonempty classes of probability laws on a common measurable space $(\mathcal X,\mathcal F)$. For a sample size $n\geq1$, let $X^n=(X_1,\ldots,X_n)$. The composite binary hypothesis testing problem is
\begin{equation*}
\begin{aligned}
H_{\mathcal C_0} &:X^n\sim P^{\otimes n}\quad\text{for some }P\in\mathcal C_0,\quad {\rm vs.} \quad H_{\mathcal C_1} &:X^n\sim Q^{\otimes n}\quad\text{for some }Q\in\mathcal C_1.
\end{aligned}
\end{equation*}
Let $\varphi_n:\mathcal X^n\to[0,1]$ be an $\mathcal F^{\otimes n}$ measurable randomised test at sample size $n$. Randomisation may be necessary to attain the Type I constraint exactly at finite sample size, since a deterministic decision rule may not achieve an arbitrary prescribed value of $\varepsilon$. Moreover, randomisation at the threshold permits this exact calibration. For an observation $x^n$, $\varphi_n(x^n)$ is the probability of deciding $H_{\mathcal C_1}$. Its Type I error over the null class is
\begin{equation*}
\alpha_n(\varphi_n;\mathcal C_0) :=\sup_{P\in\mathcal C_0}\mathbb E_{P^{\otimes n}} \left[\varphi_n(X^n)\right].
\end{equation*}
For $\varepsilon\in(0,1)$, define
\begin{equation}
\beta_n^\star(\varepsilon;\mathcal C_0,\mathcal C_1):=\inf_{\varphi_n:\,\alpha_n(\varphi_n;\mathcal C_0)\leq\varepsilon} \sup_{Q\in\mathcal C_1} \mathbb E_{Q^{\otimes n}} \left[1-\varphi_n(X^n)\right].
\label{eq:composite-minimax}
\end{equation}
This is the optimal Type II error at sample size $n$ under the uniform Type I error constraint $\varepsilon$. The infimum is taken over all randomised tests satisfying this constraint.

%##################################################
\subsection{Definitions}
% -------------------------------------------------
\begin{definition} 
Let $P$ and $Q$ be probability laws on $(\mathcal X,\mathcal F)$. For $\lambda>0$, $\lambda\neq1$, let $\mu$ be any measure with respect to which both $P$ and $Q$ admit densities. The order-$\lambda$ Hellinger integral~\cite{van2014renyi} is
\begin{equation}
H_\lambda(Q,P):=\int_{\mathcal X}\left(\frac{dQ}{d\mu}\right)^\lambda\left(\frac{dP}{d\mu}\right)^{1-\lambda}\,d\mu,
\label{eq:hellinger-integral}
\end{equation}
and the R\'enyi divergence~\cite{van2014renyi} of order $\lambda$ is
\begin{equation}
D_\lambda(Q\|P):=\frac{1}{\lambda-1}\log H_\lambda(Q,P).
\label{eq:renyi-divergence-definition}
\end{equation}
\end{definition}
These quantities do not depend on the choice of $\mu$. Standard extended real conventions apply, and $D(Q\|P)$ denotes the KL divergence.
% -------------------------------------------------
\begin{definition}
For $\lambda>0$, $\lambda\neq1$, define
\begin{equation*}
D_\lambda(\mathcal C_1\|\mathcal C_0):=\inf_{Q\in\mathcal C_1} \inf_{P\in\mathcal C_0}D_\lambda(Q\|P), \quad D(\mathcal C_1\|\mathcal C_0):= \inf_{Q\in\mathcal C_1} \inf_{P\in\mathcal C_0} D(Q\|P).
\end{equation*}
\end{definition}

\begin{definition}
For $\lambda \in (0,1)$, a pair $(Q_\lambda^\star,P_\lambda^\star) \in\mathcal C_1\times\mathcal C_0$ is called a joint R\'enyi projection if
\begin{equation*}
D_\lambda(Q_\lambda^\star\|P_\lambda^\star)=\inf_{Q\in\mathcal C_1} \inf_{P\in\mathcal C_0} D_\lambda(Q\|P)= D_\lambda(\mathcal C_1\|\mathcal C_0).
\end{equation*}
We refer to $(Q_\lambda^\star,P_\lambda^\star)$ as the \textit{projected pair}. Since $\lambda<1$, this is equivalent to $(Q_\lambda^\star,P_\lambda^\star) \in \operatorname*{arg\,max}\limits_{(Q,P)\in\mathcal C_1\times\mathcal C_0} H_\lambda(Q,P).$
\end{definition}

%###################################################
\section{Bounds}
\label{sec:finite-blocklength-bounds}

We now derive finite sample bounds on the optimal Type II error~\eqref{eq:composite-minimax} under a fixed Type I error constraint $\varepsilon$. 

%##################################################
\subsection{Converse bound}
\label{sec:composite-converse}

For the converse bound, any test admissible for the composite problem is also admissible for every simple binary problem obtained by fixing $P\in\mathcal C_0$ and $Q\in\mathcal C_1$. Applying the simple binary R\'enyi converse in \cite[Thm.~1]{bruno2026finite} to each such pair, then taking the infimum over the two classes and optimising over $\lambda>1$, gives the following composite bound.

The following theorem gives a finite sample lower bound on $\beta_n^\star(\varepsilon;\mathcal C_0,\mathcal C_1)$ using a R\'enyi divergence of order $\lambda>1$.

\begin{theorem}
\label{thm:finite-blocklength-composite-renyi-converse}
For every $\varepsilon\in(0,1)$ and $n\geq1$,
\begin{equation}
\beta_n^\star(\varepsilon;\mathcal C_0,\mathcal C_1) \geq 1- \exp\left\{-\sup_{\lambda>1} \frac{\lambda-1}{\lambda} \left[\log\frac{1}{\varepsilon} -n\inf_{Q\in\mathcal C_1}\inf_{P\in\mathcal C_0}D_\lambda(Q\|P) \right]_+ \right\},
\label{eq:finite-blocklength-composite-renyi-converse}
\end{equation}
where $[a]_+:=\max\{a,0\}$, with $[-\infty]_+:=0$.
\end{theorem}

The proof of \Cref{thm:finite-blocklength-composite-renyi-converse}
is given in \Cref{app:finite-blocklength-composite-renyi-converse}.

%###################################################
\subsection{Achievability bound}
\label{sec:composite-achievability}

Constructing an achievability bound requires a single test that controls, uniformly, the Type I error over the null class and the Type II error  over the alternative class. We first identify conditions under which a single statistic provides this uniform control.

\begin{lemma}
\label{lem:uniform-score-achievability}
Fix $n\geq1$, $\varepsilon\in(0,1)$, $\lambda\in(0,1)$, and $\xi_0,\xi_1\geq0$. Suppose that a measurable function
$h:\mathcal X\to[-\infty,+\infty]$ satisfies
\begin{equation*}
\sup_{P\in\mathcal C_0} \mathbb E_P\left[e^{\lambda h}\right] \leq e^{(\lambda-1)\xi_0}, \qquad \sup_{Q\in\mathcal C_1}\mathbb E_Q\left[e^{(\lambda-1)h}\right] \leq e^{(\lambda-1)\xi_1}.
\end{equation*}
For every
\begin{equation}
\tau \geq \frac{\log(1/\varepsilon)-n(1-\lambda)\xi_0}{\lambda},\qquad \psi_{n,\tau}(x^n)=\mathbbm 1 \left\{\sum_{i=1}^n h(x_i)\geq\tau\right\},
\label{eq:threshold-tau}
\end{equation}
satisfying $\alpha_n(\psi_{n,\tau};\mathcal C_0) \leq \varepsilon,$ and
\begin{equation*}
\beta_n(\psi_{n,\tau};\mathcal C_1):=\sup_{Q\in\mathcal C_1} \mathbb E_{Q^{\otimes n}} \left[ 1-\psi_{n,\tau}(X^n) \right] \leq \exp\left\{ (1-\lambda)\tau+n(\lambda-1)\xi_1 \right\}.
\end{equation*}
\end{lemma}
The proof of \Cref{lem:uniform-score-achievability} is given in \Cref{app:uniform-score-achievability}.

The role of \Cref{lem:uniform-score-achievability} is to isolate what is required for achievability in the composite setting. A single statistic is sufficient if its two exponential moments can be controlled uniformly over the two classes. For a fixed pair $P$ and $Q$, taking $h=\log(Q/P)$ gives $\mathbb E_P[e^{\lambda h}] =\mathbb E_Q[e^{(\lambda-1)h}] =H_\lambda(Q,P)$. These moments are therefore determined by the R\'enyi divergence through \eqref{eq:renyi-divergence-definition}. Setting the threshold to the smallest value allowed in \eqref{eq:threshold-tau} recovers the simple binary finite sample R\'enyi achievability bound in~\cite[Thm.~2]{bruno2026finite}.

The composite problem is therefore to find a single pair whose log-likelihood ratio provides this exponential control uniformly over the two classes. We select a pair that maximises the Hellinger integral over the classes. For $\lambda\in(0,1)$, this is equivalent to minimising the R\'enyi divergence, and hence to finding a joint R\'enyi projection $(Q_\lambda^\star,P_\lambda^\star)$. The convexity of the hypothesis classes, together with the concavity of the Hellinger integral, allows us to show that the log-likelihood ratio of this pair satisfies the required uniform bounds.

Throughout this section, weak compactness and weak semicontinuity in $L^1(\mu)$ are understood with respect to the weak topology $\sigma(L^1(\mu),L^\infty(\mu))$, under which $f_n\to f \iff \int f_n g\,d\mu\to\int fg\,d\mu, \ \forall g\in L^\infty(\mu)$.

\begin{theorem}
\label{thm:projected-finite-blocklength-achievability}
Fix $\lambda\in(0,1)$. Assume that all laws in $\mathcal C_0$ and $\mathcal C_1$ admit densities with respect to a common $\sigma$ finite measure $\mu$, and that the two classes are convex and weakly compact as defined above. Then a joint R\'enyi projection $(Q_\lambda^\star,P_\lambda^\star)$ exists. If one can be chosen such that $H_\lambda(Q_\lambda^\star,P_\lambda^\star)>0$ and $R\ll P_\lambda^\star+Q_\lambda^\star$ for all $R\in\mathcal C_0\cup\mathcal C_1$, then, for every $n\geq1$ and $\varepsilon\in(0,1)$,
\begin{equation}
\begin{aligned}
\beta_n^\star(\varepsilon;\mathcal C_0,\mathcal C_1)
\leq \min\Bigg\{1-\varepsilon, \exp\left[-\frac{1-\lambda}{\lambda}\left(n\inf_{Q\in\mathcal C_1}\inf_{P\in\mathcal C_0}D_\lambda(Q\|P)-\log\frac{1}{\varepsilon}\right)\right]\Bigg\}.
\end{aligned}
\label{eq:projected-finite-blocklength-achievability}
\end{equation}
Moreover, the exponential upper bound is obtained by a threshold test constructed from the joint R\'enyi projection, with threshold
\begin{equation}
\tau=\frac{\log(1/\varepsilon)-n(1-\lambda)
\inf_{Q\in\mathcal C_1}\inf_{P\in\mathcal C_0}
D_\lambda(Q\|P)}{\lambda}.
\label{eq:proj-threshold}
\end{equation}
\end{theorem}

The proof of \Cref{thm:projected-finite-blocklength-achievability} is given in \Cref{app:projected-finite-blocklength-achievability}.

\Cref{thm:projected-finite-blocklength-achievability} gives the finite sample achievability bound~\eqref{eq:projected-finite-blocklength-achievability} without assuming that the projected pair is least favourable. 

%###################################################
\subsection{Phase transition under an exponential Type I constraint}
\label{subsec:threshold-rate}
We now specialise the finite sample bounds to a Type I error probability constraint that decays exponentially with sample size. For $r>0$, set $\varepsilon=e^{-nr}.$

\begin{theorem}
\label{cor:threshold-rate}
Fix $r>0$ and assume that
\begin{equation*} 
\lim_{\lambda\uparrow1}D_\lambda(\mathcal C_1\|\mathcal C_0) = \lim_{\lambda\downarrow1}D_\lambda(\mathcal C_1\|\mathcal C_0) = D(\mathcal C_1\|\mathcal C_0).
\end{equation*} 
Suppose moreover that there exists a sequence $\lambda_k\uparrow1$, with $\lambda_k\in(0,1)$, such that the assumptions of \Cref{thm:projected-finite-blocklength-achievability} hold at every $\lambda_k$. Then
\begin{equation} 
\lim_{n\to\infty}\beta_n^\star(e^{-nr};\mathcal C_0,\mathcal C_1) = \begin{cases} 0, & 0<r<D(\mathcal C_1\|\mathcal C_0),\\ 1, & r>D(\mathcal C_1\|\mathcal C_0).
\end{cases}
\label{eq:threshold-KL}
\end{equation}
No assertion is made when $r=D(\mathcal C_1\|\mathcal C_0)$.
\end{theorem}
The proof of \Cref{cor:threshold-rate} is given in \Cref{app:threshold-rate}.

The order one limits assumed in \Cref{cor:threshold-rate} do not follow directly from the corresponding fixed pair result because the pair minimising the R\'enyi divergence may vary with the order. The following lemma gives a sufficient condition for convergence from above, while \Cref{cor:composite-renyi-order-one-below} (\Cref{app:composite-renyi-order-one-below}) gives one for convergence from below.

\begin{lemma}
\label{lem:composite-renyi-order-one}
Assume that $D(\mathcal C_1\|\mathcal C_0)<+\infty$ and that, for every $\eta>0$, there exist $P_\eta\in\mathcal C_0$, $Q_\eta\in\mathcal C_1$, and $\delta_\eta>0$ such that $D(Q_\eta\|P_\eta) \leq D(\mathcal C_1\|\mathcal C_0)+\eta $ and $D_{1+\delta_\eta}(Q_\eta\|P_\eta)<+\infty.$ Then
\begin{equation*}
\lim_{\lambda\downarrow1} D_\lambda(\mathcal C_1\|\mathcal C_0) = D(\mathcal C_1\|\mathcal C_0).
\end{equation*}
\end{lemma}
The proof of \Cref{lem:composite-renyi-order-one} is given in \Cref{app:composite-renyi-order-one}.

For simple binary testing, the finite sample achievability and converse bounds involve R\'enyi orders below and above one, respectively\cite{bruno2026finite}. Their limits as the order approaches one recover the KL divergence and hence the phase transition threshold. In the composite setting, the same mechanism applies only after establishing the corresponding order-one limits for $D_\lambda(\mathcal C_1\|\mathcal C_0)$, since the pair minimising the R\'enyi divergence may depend on $\lambda$.

We now specialise the result to compact convex classes with full support on a finite alphabet and characterise the exact error exponents on both sides of the phase transition.

%####################################
%####################################
\section{Exact error exponents on finite alphabets}
\label{sec:finite-alphabet-consequences}

Let $\mathcal X=\{1,\ldots,d\}$ be a finite alphabet and let
$\Delta_d:=\left\{p\in[0,1]^d \mathrel{}\middle|\mathrel{} \sum_{x=1}^d p(x)=1\right\}$
denote the probability simplex on $\mathcal X$. Throughout this section,
$\mathcal C_0,\mathcal C_1\subseteq\Delta_d$ are nonempty, compact, and convex, and every distribution in $\mathcal C_0\cup\mathcal C_1$ has full support.

Compactness and full support give a common strictly positive lower bound on the probability assigned to every point in the alphabet by every distribution in the two classes. Thus all distributions are mutually absolutely continuous, every pair has a strictly positive Hellinger integral, and the support conditions in \Cref{thm:projected-finite-blocklength-achievability} are automatic. Consequently, for every $\lambda\in(0,1)$ a projected pair
$(Q_\lambda^\star,P_\lambda^\star)$ exists, with $h_\lambda^\star(x)=\log\frac{Q_\lambda^\star(x)}{P_\lambda^\star(x)},$
and \Cref{thm:projected-finite-blocklength-achievability} applies
without further support assumptions.

Under the finite alphabet assumptions, both order one limits required by \Cref{cor:threshold-rate} hold. Convergence from below follows from \Cref{cor:composite-renyi-order-one-below} (\Cref{app:composite-renyi-order-one-below}). For convergence from above, uniform full support ensures that the R\'enyi divergence of every finite order is finite for every pair, so \Cref{lem:composite-renyi-order-one} applies. Together with the achievability conditions established above, \Cref{cor:threshold-rate} therefore identifies $D(\mathcal C_1\|\mathcal C_0)$ as the phase transition threshold.

%####################################
\subsection{Exact exponent in the achievable regime}
\label{sec:exact-exponent}

In the achievable regime
$0<r<D(\mathcal C_1\|\mathcal C_0)$, the projected test gives an achievable Type II error exponent. Every test admissible for the composite problem is also admissible for each simple binary problem obtained by selecting one distribution from each class. The composite exponent therefore cannot exceed the exponent for any fixed pair. To establish optimality, it is enough to show that the projected achievability exponent coincides with the smallest exponent over all such
pairs.

Under the finite alphabet assumptions, the optimisation over the
R\'enyi order and the pair of distributions has the required
convexity-concavity structure for Sion's minimax Theorem~\cite[Th.~3.4]{sion1958general}, which allows the two optimisations to be interchanged. Compactness then gives attainment of both extrema. This identifies an order and a projected pair for which the composite exponent coincides with the corresponding simple binary exponent.

\begin{theorem}
\label{thm:exact-composite-exponent}
Fix $0<r<D(\mathcal C_1\|\mathcal C_0)$. Then the exact Type II error
exponent exists and satisfies
\begin{align}
\lim_{n\to\infty}
-\frac{1}{n}\log \beta_n^\star(e^{-nr};\mathcal C_0,\mathcal C_1) & =\max_{0<\lambda<1} \frac{1-\lambda}{\lambda} \left[D_\lambda(\mathcal C_1\|\mathcal C_0)-r \right] \nonumber \\ &= \min_{\substack{P\in\mathcal C_0\\Q\in\mathcal C_1}} \max_{0<\lambda<1}\frac{1-\lambda}{\lambda}\left[D_\lambda(Q\|P)-r\right].
\label{eq:exact-composite-exponent}
\end{align}
Moreover, there exist $\lambda_r^\star\in(0,1)$,$P_r^\star\in\mathcal C_0$, and $Q_r^\star\in\mathcal C_1$ such that $(Q_r^\star,P_r^\star)$ attains $D_{\lambda_r^\star}(\mathcal C_1\|\mathcal C_0)$ and $\lambda_r^\star$ attains the inner maximum for $(P_r^\star,Q_r^\star)$. The common value is
\begin{equation*}
\frac{1-\lambda_r^\star}{\lambda_r^\star}\left[D_{\lambda_r^\star}(Q_r^\star\|P_r^\star)-r\right].
\end{equation*}
\end{theorem}
\Cref{eq:exact-composite-exponent} shows that, among all simple pairs $(P,Q)\in\mathcal C_0\times\mathcal C_1$, the pair $(P_r^\star,Q_r^\star)$ attains the smallest Type II error exponent. The latter, is also the exponent of the composite problem, so $(P_r^\star,Q_r^\star)$ determines the decay rate of the optimal composite Type II error. In other words, $(P_r^\star,Q_r^\star)$ is least favourable at the level of the error exponent.

\begin{corollary}
\label{cor:hoeffding-tangent}
For every $0<r<D(\mathcal C_1\|\mathcal C_0)$, the R\'enyi order $\lambda_r^\star$ in \Cref{thm:exact-composite-exponent} is unique. Moreover, the exact Type II error exponent is continuously differentiable with respect to $r$, with derivative: $-(1-\lambda_r^\star)/\lambda_r^\star.$
\end{corollary}
The proof is given in \Cref{app:hoeffding-tangent}.

\Cref{cor:hoeffding-tangent} shows that the exact achievable exponent is attained at a unique R\'enyi order, and that this order can be recovered directly from the derivative of the exponent.

%####################################
\subsection{Exact exponent in the converse regime}
\label{sec:exact-strong-converse-exponent}

We now consider $r>D(\mathcal C_1\|\mathcal C_0)$, where \Cref{cor:threshold-rate} shows that the optimal Type II error converges to one. The nontrivial quantity is $1-\beta_n^\star(e^{-nr};\mathcal C_0,\mathcal C_1)$. The finite sample converse in \Cref{thm:finite-blocklength-composite-renyi-converse} provides an exponential upper bound on this quantity. The following theorem determines its decay exponent and shows that the exponent in the converse is attained.

\begin{theorem}
\label{thm:exact-composite-strong-converse-exponent}
For every $r>D(\mathcal C_1\|\mathcal C_0)$, the exact exponent of the
residual probability exists and satisfies
\begin{align}
\lim_{n\to\infty}-\frac{1}{n}\log\left(1-\beta_n^\star(e^{-nr};\mathcal C_0,\mathcal C_1)\right)&=\sup_{\lambda>1}\frac{\lambda-1}{\lambda}\left[r-D_\lambda(\mathcal C_1\|\mathcal C_0)\right]_+
\label{eq:exact-composite-strong-converse-exponent}
\\&=\sup_{\substack{P\in\mathcal C_0\\Q\in\mathcal C_1}}\sup_{\lambda>1}\frac{\lambda-1}{\lambda}\left[r-D_\lambda(Q\|P)\right]_+.
\label{eq:composite-strong-converse-pairwise}
\end{align}
Consequently,
\begin{equation}
1-\beta_n^\star(e^{-nr};\mathcal C_0,\mathcal C_1)=\exp\left\{-n\sup_{\substack{P\in\mathcal C_0\\Q\in\mathcal C_1}}\sup_{\lambda>1} \frac{\lambda-1}{\lambda} \left[r-D_\lambda(Q\|P)\right]_+ +o(n) \right\}.
\label{eq:strong-converse-exact-asymptotic}
\end{equation}
\end{theorem}

The proof is given in \Cref{app:exact-composite-strong-converse-exponent}.

\Cref{thm:exact-composite-strong-converse-exponent} completes the exponent level characterisation on both sides of the phase transition threshold $D(\mathcal C_1\|\mathcal C_0)$ identified in \Cref{cor:threshold-rate}. In the achievable regime, the exact Type II error exponent is the smallest exponent among the corresponding simple binary problems. In the converse regime, the exponent governing convergence of the Type II error to one is the largest strong converse exponent among the simple pairs.

%####################################
\subsection{Fixed Type I constraint and zero rate limit}
\label{sec:fixed-type-one}

The exponentially decaying Type I constraint excludes the boundary case $r=0$. We now consider this boundary by fixing $\varepsilon\in(0,1)$. For simple binary testing, the Chernoff-Stein lemma gives the KL divergence from the null to the alternative, reversing the direction that determines the threshold rate. In the projected achievability bound, the relevant R\'enyi orders approach zero as $n\to\infty$. We first determine the corresponding order zero limit.

\begin{lemma}
\label{lem:order-zero-limit}
For every $P\in\mathcal C_0$, $Q\in\mathcal C_1$, and $\lambda\in(0,1)$,
\begin{equation*}
D(P\|Q)-\frac{\lambda}{2} \left[\log\frac{1}{\min_{\substack{R\in\mathcal C_0\cup\mathcal C_1\\ x\in\mathcal X}} R(x)}\right]^2 \leq\frac{1-\lambda}{\lambda} D_\lambda(Q\|P) \leq D(P\|Q).
\end{equation*}
Consequently,
\begin{equation}
\lim_{\lambda\downarrow0} \frac{1-\lambda}{\lambda} D_\lambda(\mathcal C_1\|\mathcal C_0)= D(\mathcal C_0\|\mathcal C_1).
\label{eq:order-zero-limit}
\end{equation}
\end{lemma}
The proof is given in \Cref{app:order-zero-limit}.

The fixed Type I exponent for compact convex classes on a finite
alphabet is a known composite Chernoff-Stein result \cite{brandao2020adversarial,mosonyi2021error}. We recover it here directly from the finite sample R\'enyi achievability bound. This also identifies the fixed Type I regime with the order-zero boundary of the R\'enyi expression used above.
\begin{corollary}
\label{thm:fixed-type-one-exponent}
For every fixed $\varepsilon\in(0,1)$,
\begin{equation}
\lim_{n\to\infty}-\frac{1}{n} \log \beta_n^\star(\varepsilon;\mathcal C_0,\mathcal C_1)=D(\mathcal C_0\|\mathcal C_1).
\label{eq:fixed-type-one-exponent}
\end{equation}
If $D(\mathcal C_0\|\mathcal C_1)=0$, then 
\begin{equation*}
\beta_n^\star(\varepsilon;\mathcal C_0,\mathcal C_1)=1-\varepsilon \qquad \forall n\geq1.
\end{equation*}
\end{corollary}
The proof is given in \Cref{app:fixed-type-one-exponent}.

The fixed Type I constraint and the exponentially decaying Type I
constraint lead to opposite KL directions. The phase transition threshold is $D(\mathcal C_1\|\mathcal C_0)$, while under a fixed Type I constraint $\varepsilon\in(0,1)$ the Type II error exponent is $D(\mathcal C_0\|\mathcal C_1)$. The former arises from R\'enyi orders approaching one, while the latter arises from orders approaching zero. When $\varepsilon=e^{-nr}$, the achievable exponent converges to the fixed Type I exponent as $r\downarrow0$.
\begin{corollary}
\label{cor:zero-rate-boundary}
If $D(\mathcal C_0\|\mathcal C_1)>0$, then $\lim_{r\downarrow0} \max_{0<\lambda<1} \frac{1-\lambda}{\lambda} \left[ D_\lambda(\mathcal C_1\|\mathcal C_0)-r \right]=D(\mathcal C_0\|\mathcal C_1).$
\end{corollary}
The proof is given in \Cref{app:zero-rate-boundary}.

%###################################################
\section{Refining the analysis}
\label{sec:finite-blocklength-exactness}

Throughout this section, we retain the finite alphabet assumptions of \Cref{sec:finite-alphabet-consequences}. We first keep the projected log-likelihood ratio~\Cref{eq:projected-log-likelihood-ratio} fixed and optimise its threshold directly under an arbitrary composite Type I constraint $\varepsilon\in(0,1)$. We then return to the exponentially decaying constraint $\varepsilon=e^{-nr}$ in the achievable regime $0<r<D(\mathcal C_1\|\mathcal C_0)$, where \Cref{thm:exact-composite-exponent} identifies the exact Type II error exponent. In this regime, we determine the polynomial order of the optimal Type II error and the corresponding logarithmic correction to the threshold
in~\eqref{eq:proj-threshold} in \Cref{thm:projected-finite-blocklength-achievability}.

The preceding results are sufficient to recover the correct exponential behaviour of the optimal Type II error and the phase transition between the achievable and converse regimes. An advantage of the achievability bound in \Cref{thm:projected-finite-blocklength-achievability} is that it is explicit and amenable to analytical study. It can, however, be tightened at finite sample size at the cost of a less explicit construction that, in general, requires numerical evaluation.
%###################################################
\subsection{Tightening the achievability bound}
\label{subsec:projected-tightening}

The threshold in~\eqref{eq:proj-threshold} is chosen using an exponential upper bound on the Type I error so that the constraint holds uniformly for every $P\in\mathcal C_0$. Consequently, the resulting Type I error may be strictly smaller than $\varepsilon$. A tighter bound is obtained by keeping the log-likelihood ratio selected by the R\'enyi projection fixed and calibrating its threshold directly against the composite Type I constraint.

\begin{proposition}
\label{prop:projected-tightening} 
Fix $\lambda\in(0,1)$ and a projected pair $(Q_\lambda^\star,P_\lambda^\star)$ satisfying \Cref{thm:projected-finite-blocklength-achievability}, and define 
\begin{equation} 
h_\lambda^\star(x) := \log\frac{Q_\lambda^\star(x)}{P_\lambda^\star(x)}.
\label{eq:projected-log-likelihood-ratio}
\end{equation} 
For each $n\geq1$, $\tau\in\mathbb R$, and $\eta\in[0,1]$, define 
\begin{equation*} \psi_{\tau,\eta}^\star(x^n) := \mathbbm 1 \left\{ \sum_{i=1}^n h_\lambda^\star(x_i)>\tau \right\} + \eta\mathbbm 1 \left\{ \sum_{i=1}^n h_\lambda^\star(x_i)=\tau \right\}. 
\end{equation*} 
Then, for every $n\geq1$ and $\varepsilon\in(0,1)$, there exist $\widehat\tau\in\mathbb R$ and $\widehat\eta\in[0,1]$ such that $\sup\limits_{P\in\mathcal C_0} \mathbb E_{P^{\otimes n}} \left[ \psi_{\widehat\tau,\widehat\eta}^\star \right]\hspace{-0.1cm}= \varepsilon.$ The induced test uniquely minimises the Type II error over all tests $\psi_{\tau,\eta}^\star$ satisfying the Type I constraint. Moreover, 
\begin{equation} \beta_n^\star(\varepsilon;\mathcal C_0,\mathcal C_1) \leq \sup_{Q\in\mathcal C_1} \mathbb E_{Q^{\otimes n}} \left[ 1-\psi_{\widehat\tau,\widehat\eta}^\star \right].
\label{eq:projected-threshold-type-two-value}
\end{equation}
\end{proposition}
The proof of \Cref{prop:projected-tightening} is given in \Cref{app:projected-tightening}.

For a fixed R\'enyi order, \Cref{prop:projected-tightening} gives the smallest Type II upper bound obtainable from threshold tests based on the corresponding projected log-likelihood ratio~\Cref{eq:projected-log-likelihood-ratio}. In particular, it cannot be worse than the explicit threshold choice in \Cref{thm:projected-finite-blocklength-achievability}. The bound may be tightened further by optimising over the R\'enyi order, but the projected pair and calibrated threshold then both depend on the order. This optimisation is generally less amenable to analytical treatment, and in \Cref{sec:numerical-illustrations} we evaluate it numerically.

The uniqueness in \Cref{prop:projected-tightening} concerns the induced test function, not necessarily its representation by $(\tau,\eta)$.Different pairs $(\tau,\eta)$ can represent the same boundary rule.For singleton classes, \Cref{app:simple-binary-threshold-tightening} shows that $h_\lambda^\star$ is the ordinary log-likelihood ratio and the optimised threshold test is the NP test. For general composite classes, the optimisation remains restricted to threshold tests based on a projected log-likelihood ratio~\Cref{eq:projected-log-likelihood-ratio} and is not asserted to solve the unrestricted composite problem.

%###################################################
\subsection{Polynomial refinement}
\label{subsec:polynomial-refinement}

\Cref{thm:exact-composite-exponent} determines the exponential rate, while \Cref{prop:projected-tightening} shows that the projected threshold may be improved when the projected log-likelihood ratio~\Cref{eq:projected-log-likelihood-ratio} (\cref{eq:projected-log-likelihood-ratio}) is held fixed. We now determine the behaviour beyond the exponential scale at the selected order $\lambda_r^\star$. Under the exponential constraint $\varepsilon=e^{-nr}$, the threshold ($\tau$) given by \Cref{thm:projected-finite-blocklength-achievability} at the selected order is
\begin{equation*}
n\left[r-\frac{1-\lambda_r^\star}{\lambda_r^\star}\left(D_{\lambda_r^\star}(Q_r^\star\|P_r^\star)-r\right)\right].
\end{equation*}
A sharper analysis of the Type I error at  reveals an additional factor of order $n^{-1/2}$. The projected threshold ($\tau$) therefore leaves part of the available Type I error unused. Recovering it requires a correction of order $\log n$ and produces a polynomial factor in the Type II error.

\begin{proposition}
\label{prop:polynomial-refinement}
Fix $0<r<D(\mathcal C_1\|\mathcal C_0)$, and let $(\lambda_r^\star,P_r^\star,Q_r^\star)$ satisfy \Cref{thm:exact-composite-exponent}. Then
\begin{equation}
\beta_n^\star(e^{-nr};\mathcal C_0,\mathcal C_1) =\Theta\!\left(n^{-1/(2\lambda_r^\star)}\exp\left\{-n\frac{1-\lambda_r^\star}{\lambda_r^\star}\left[D_{\lambda_r^\star}(Q_r^\star\|P_r^\star)-r\right]\right\}\right).
\label{eq:polynomial-refinement}
\end{equation}
Moreover, the matching upper bound is achieved by a projected log-likelihood ratio~\Cref{eq:projected-log-likelihood-ratio} test with threshold
\begin{equation*}
n\left[r-\frac{1-\lambda_r^\star}{\lambda_r^\star}\left(D_{\lambda_r^\star}(Q_r^\star\|P_r^\star)-r\right)\right]-\frac{\log n}{2\lambda_r^\star}+O(1).
\end{equation*}
\end{proposition}
The proof is given in \Cref{app:polynomial-refinement}.

Equation~\eqref{eq:polynomial-refinement} determines the polynomial order but does not identify the leading multiplicative constant under the assumptions of \Cref{prop:polynomial-refinement}. Stronger constant level conclusions require additional regularity conditions.

\begin{corollary}
\label{cor:polynomial-slope}
For every $0<r<D(\mathcal C_1\|\mathcal C_0)$, the polynomial power in~\eqref{eq:polynomial-refinement} satisfies
\begin{equation}
\frac{1}{2\lambda_r^\star}=\frac{1}{2}\left[1-\frac{d}{dr}\max_{0<\lambda<1}\frac{1-\lambda}{\lambda}\left[D_\lambda(\mathcal C_1\|\mathcal C_0)-r\right]\right].
\label{eq:hoeffding-tangent-polynomial}
\end{equation}
\end{corollary}
The proof of \Cref{cor:polynomial-slope} is given in
\Cref{app:polynomial-slope}.

\Cref{cor:polynomial-slope} shows that the polynomial power is determined directly by the slope of the exact Type II error exponent. The associated projected pair $(Q_r^\star,P_r^\star)$, however, need not be unique.

%###########################################
\section{When is the projected pair least favourable?}
\label{sec:exact-reduction}

Classical robust testing results establish least favourability under structural conditions such as stochastic ordering or related dominance properties of the likelihood ratio \cite{huber1965robust,verdu2003minimax,fauss2021minimax}. Under such conditions, a particular pair of distributions attains the suprema defining the Type I and Type II errors, reducing the composite problem to a simple binary test.
 
The R\'enyi projection used in the preceding sections is selected because it provides uniform exponential control over the two classes. This property alone does not imply that the projected pair is least favourable at finite sample size. G\"ul~\cite{gul2026structural} gives convex uncertainty classes for which a single pair is optimal for every R\'enyi order in $(0,1)$ but fails the stronger convex order condition required for finite sample least favourability.
 
We therefore ask under what additional conditions the projected pair also attains the composite error probabilities. Fix $\lambda\in(0,1)$ and a projected pair $(Q_\lambda^\star,P_\lambda^\star)$, and let $\psi_{\tau,\eta}^\star$ denote the projected threshold test introduced in \Cref{subsec:projected-tightening}. The following condition imposes stochastic ordering directly on the projected log-likelihood ratio~\Cref{eq:projected-log-likelihood-ratio}.

%--------------------------------------------------- 
\begin{proposition} 
\label{prop:projected-ordering} 
Suppose that, for every $t\in\mathbb R$, $\forall P\in\mathcal C_0,$ and $ \forall Q\in\mathcal C_1,$ 
\begin{equation*} P(h_\lambda^\star\geq t) \leq P_\lambda^\star(h_\lambda^\star\geq t), \quad {\rm and} \quad Q(h_\lambda^\star\leq t) \leq Q_\lambda^\star(h_\lambda^\star\leq t). \end{equation*} 
Then, for every $n\geq1$, $\tau\in\mathbb R$, and $\eta\in[0,1]$, 
\begin{equation*} \sup_{P\in\mathcal C_0} \mathbb E_{P^{\otimes n}}[\psi_{\tau,\eta}^\star] = \mathbb E_{(P_\lambda^\star)^{\otimes n}} [\psi_{\tau,\eta}^\star], \quad {\rm and} \quad \sup_{Q\in\mathcal C_1} \mathbb E_{Q^{\otimes n}}[1-\psi_{\tau,\eta}^\star] = \mathbb E_{(Q_\lambda^\star)^{\otimes n}} [1-\psi_{\tau,\eta}^\star]. 
\end{equation*} 
\end{proposition}
The proof is given in \Cref{app:projected-ordering}.
 
Thus, under the stochastic ordering conditions of \Cref{prop:projected-ordering}, the projected pair attains both composite error suprema for every projected threshold test. To obtain an exact reduction of the unrestricted composite problem, the same threshold test must also be optimal for the projected simple pair.

%---------------------------------------------------
\begin{corollary}
\label{cor:projected-exact-reduction}
Assume the conditions of \Cref{prop:projected-ordering}. Fix $n\geq1$ and $\varepsilon\in(0,1)$. Suppose that $\psi_{\tau,\eta}^\star$ is optimal for testing $(P_\lambda^\star)^{\otimes n}$ against $(Q_\lambda^\star)^{\otimes n}$ under the constraint $\mathbb E_{(P_\lambda^\star)^{\otimes n}}[\psi_{\tau,\eta}^\star]\leq\varepsilon.$ Then
\begin{equation*}
\beta_n^\star(\varepsilon;\mathcal C_0,\mathcal C_1)=\beta_n^\star(\varepsilon;P_\lambda^\star,Q_\lambda^\star).
\end{equation*}
\end{corollary}
The proof is given in \Cref{app:projected-exact-reduction}.

The stochastic ordering conditions in
\Cref{prop:projected-ordering} hold for separated one parameter natural
exponential families.

\begin{definition}
Let $\{P_\theta:\theta\in\Theta\}$ have common support and densities $p_\theta(x) = h_0(x)\exp\{\theta T(x)-\psi(\theta)\},$ and $\theta\in\Theta,$ where $\Theta\subseteq\mathbb R$ is an open interval, $h_0>0$, and $\psi$ is finite, differentiable, and strictly convex. Let $\mathcal C_0$ and $\mathcal C_1$ consist of $P_\theta$ over two disjoint closed intervals
$[\theta_-^{\mathcal C_0},\theta_+^{\mathcal C_0}]$ and
$[\theta_-^{\mathcal C_1},\theta_+^{\mathcal C_1}]$, respectively.
\end{definition}

%---------------------------------------------------
\begin{proposition}
\label{prop:endpoint-reduction}
For every $n\geq1$ and $\varepsilon\in(0,1)$,
\begin{equation*} 
\beta_n^\star(\varepsilon;\mathcal C_0,\mathcal C_1) = 
\begin{cases} \beta_n^\star\left( \varepsilon; P_{\theta_+^{\mathcal C_0}}, P_{\theta_-^{\mathcal C_1}} \right), & \theta_+^{\mathcal C_0}<\theta_-^{\mathcal C_1}, \\[1ex] \beta_n^\star\left( \varepsilon; P_{\theta_-^{\mathcal C_0}}, P_{\theta_+^{\mathcal C_1}} \right), & \theta_+^{\mathcal C_1}<\theta_-^{\mathcal C_0}. 
\end{cases}
\end{equation*}
Moreover, for every $\lambda\in(0,1)$, the corresponding endpoint pairs a joint R\'enyi projection.
\end{proposition}
The proof is given in \Cref{app:endpoint-reduction}.

Hence, for separated one parameter natural exponential families, the same endpoint pair is selected by the R\'enyi projection and is least favourable at every sample size.
 
%##########################################
\section{Numerical illustrations}
\label{sec:numerical-illustrations}

Beyond characterising the phase transition threshold~\Cref{eq:threshold-KL}, and the corresponding error exponents, we illustrate the finite sample behaviour of the bounds derived above in three settings. 

First, we consider nonordered affine classes of ternary distributions. 

Second, we consider separated one parameter families, including Bernoulli, Poisson, Gaussian, and exponential models, for which~\Cref{prop:endpoint-reduction} gives an exact reduction to a simple binary testing problem. 

Third, we consider nonordered affine ternary classes under fixed and subexponentially decreasing Type I constraints, with $\varepsilon_n=0.01$ and $\varepsilon_n=1/n$. In each setting, we compare the relevant R\'enyi bounds with the optimal Type II error. 

Further details of the numerical calculations, including the evaluation of the composite R\'enyi divergences, the computation of the phase transition threshold, and the optimisation over the R\'enyi order, are provided in~\Cref{app:numerical-renyi}.

The first setting considers two affine classes of ternary distributions, for $s,t\in[0,1]$
\begin{equation*}
\begin{aligned}
P_s&=(1-s)(0.327,0.418,0.255) +s(0.563,0.266,0.171),\\
Q_t&=(1-t)(0.143,0.357,0.500) +t(0.379,0.205,0.416).
\end{aligned}
\end{equation*}
These classes are compact, convex, and have uniform full support, with $D(\mathcal C_1\|\mathcal C_0)=0.094$; numerically. 

\begin{figure}[t]
\centering
\begin{minipage}[t]{0.48\columnwidth}
    \centering
    \includegraphics[width=\linewidth]{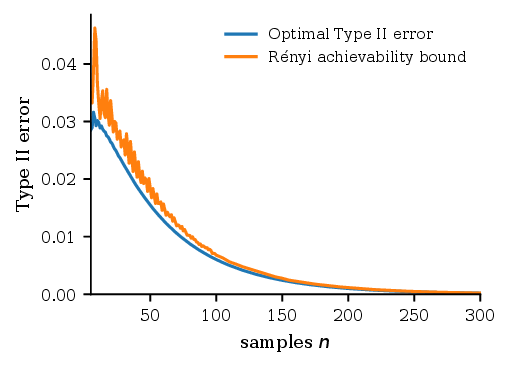}
\end{minipage}
\hfill
\begin{minipage}[t]{0.48\columnwidth}
    \centering
    \includegraphics[width=\linewidth]{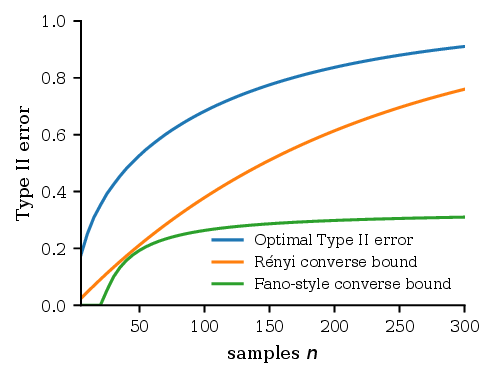}
\end{minipage}
\caption{Finite sample Type II error and R\'enyi bounds for the nonordered affine ternary classes under $\varepsilon=e^{-nr}$. The left panel considers $r=0.35D(\mathcal C_1\|\mathcal C_0)$ and the right panel considers $r=1.5D(\mathcal C_1\|\mathcal C_0)$.}
\label{fig:numerical-bounds}
\end{figure}

For the achievable regime, we take $r=0.35D(\mathcal C_1\|\mathcal C_0)=0.033$. For each sample size $n$, we compare the optimal Type II error with the achievability bound obtained by calibrating the threshold and boundary randomisation of the projected log-likelihood ratio~\Cref{eq:projected-log-likelihood-ratio} directly under the Type I constraint $\varepsilon=e^{-nr}$. The left panel of \Cref{fig:numerical-bounds} shows this comparison.

The oscillations visible in the achievability curve at small and moderate sample sizes arise from the finite number of values taken by the projected log-likelihood ratio~\Cref{eq:projected-log-likelihood-ratio} at each sample size $n$. Consequently, the admissible threshold values and the corresponding boundary randomisation change discretely with $n$.

For the converse regime, we take $r=1.5D(\mathcal C_1\|\mathcal C_0)=0.142$. The right panel of \Cref{fig:numerical-bounds} compares the R\'enyi converse in \Cref{thm:finite-blocklength-composite-renyi-converse} with the optimal Type II error. For reference, we also include a Fano-style converse bound. The R\'enyi converse is stronger over the displayed sample sizes and captures the convergence of the Type II error towards one, whereas the Fano-style converse bound remains separated from one.

The second setting considers separated one parameter Bernoulli, Poisson, Gaussian, and exponential families covered by \Cref{prop:endpoint-reduction}, under the exponentially decaying Type I constraint $\varepsilon=e^{-nr}$. For each family, we show the achievable regime at $r=0.75r_{\rm c}$ and the converse regime at $r=1.90r_{\rm c}$.

\begin{figure}[t]
\centering
\includegraphics[width=1\columnwidth]{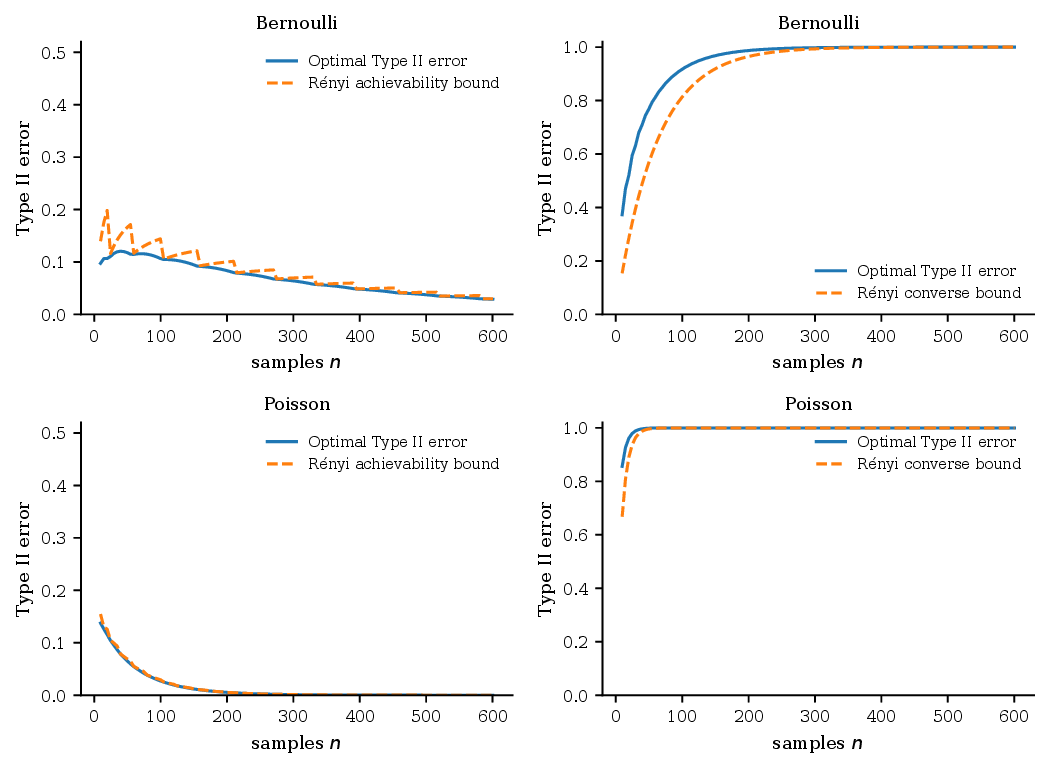}
\caption{Finite sample Type II error for the separated Bernoulli and Poisson families. The left column shows the achievable regime at $r=0.75 r_{\rm c}$, comparing the optimal Type II error with the R\'enyi achievability bound. The right column shows the converse regime at $r=1.90r_{\rm c}$, comparing the optimal Type II error with the R\'enyi converse bound.}
\label{fig:ordered-bernoulli-poisson}
\end{figure}

\begin{figure}[t]
\centering
\includegraphics[width=1\columnwidth]{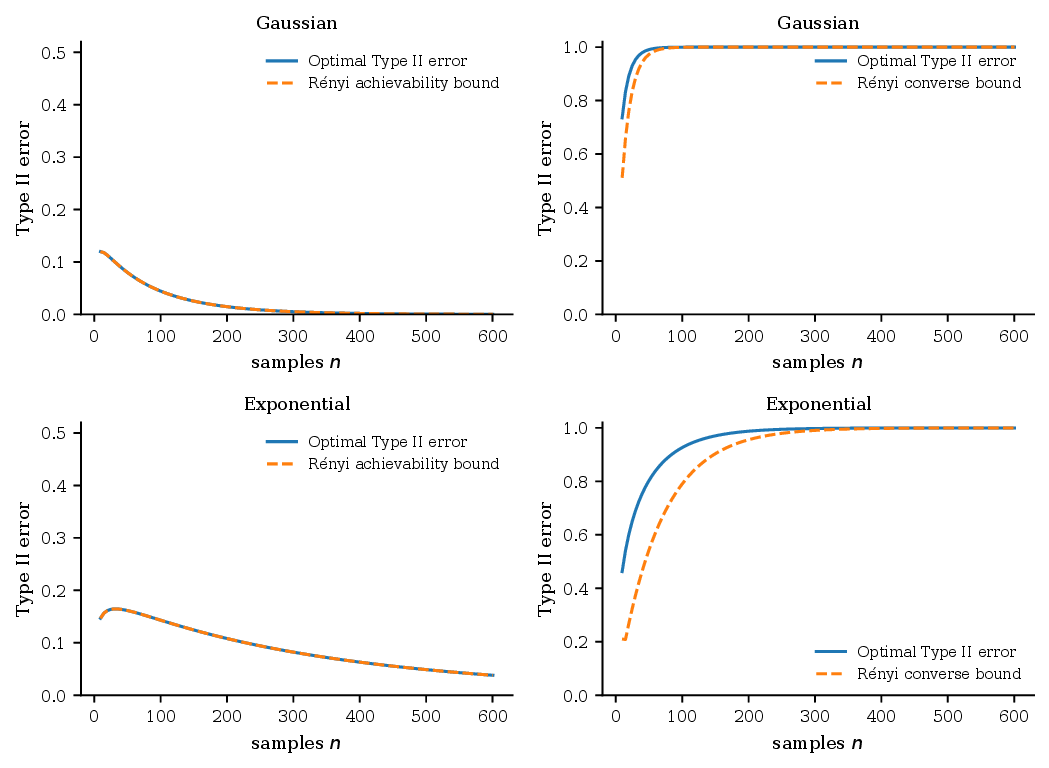}
\caption {Finite sample Type II error for the separated Gaussian and exponential families. The left column shows the achievable regime at $r=0.75r_{\rm c}$, comparing the optimal Type II error with the R\'enyi achievability bound. The right column shows the converse regime at $r=1.90r_{\rm c}$, comparing the optimal Type II error with the R\'enyi converse bound.}
\label{fig:ordered-gaussian-exponential}
\end{figure}
 
In the achievable regime, the endpoint pair selected by the R\'enyi projection is also least favourable by \Cref{prop:endpoint-reduction}, so the composite problem reduces exactly to the corresponding simple binary problem. The optimised projected threshold test therefore coincides with the NP test, and its Type II error equals the exact composite Type II error. This agreement is shown in the left columns of \Cref{fig:ordered-bernoulli-poisson,fig:ordered-gaussian-exponential}.

The converse regime behaves differently. Although~\Cref{prop:endpoint-reduction} still reduces the composite problem exactly to the corresponding simple binary problem, the finite sample R\'enyi converse gives a lower bound on the optimal Type II error rather than the exact NP performance~\cite{bruno2026finite}. The curves are therefore not expected to coincide at finite sample size. Nevertheless,the R\'enyi converse remains tight over the displayed finite sample range and closely tracks the Type II error as it approaches one,consistent with\Cref{thm:exact-composite-strong-converse-exponent}. This behaviour is shown in the right columns of\Cref{fig:ordered-bernoulli-poisson,fig:ordered-gaussian-exponential}. Consistent with the phase transition in \Cref{cor:threshold-rate}, the Type II error decreases towards zero in the left columns and approaches one in the right columns of \Cref{fig:ordered-bernoulli-poisson,fig:ordered-gaussian-exponential}.

The final setting considers composite testing when the Type I error constraint is fixed ($\varepsilon_n=0.01$) and when it decreases subexponentially with sample size ($\varepsilon_n=1/n$).

For $s,t\in[0,1]$, let
\begin{equation*}
\begin{aligned} 
\widetilde P_s&=(1-s)(0.33,0.33,0.34) +s(0.33,0.35,0.32),\\ \widetilde Q_t&=(1-t)(0.203,0.428,0.369) +t(0.370,0.455,0.175).
\end{aligned}
\end{equation*} 

Here, $\widetilde{\mathcal C}_0=\{\widetilde P_s:s\in[0,1]\}$ and $\widetilde{\mathcal C}_1=\{\widetilde Q_t:t\in[0,1]\}$. These classes are compact, convex, and have uniform full support, with $D(\widetilde{\mathcal C}_1\|\widetilde{\mathcal C}_0)=0.017$ and $D(\widetilde{\mathcal C}_0\|\widetilde{\mathcal C}_1)=0.016$; numerically.

\begin{figure}[t]
\centering
\includegraphics[width=1\columnwidth]{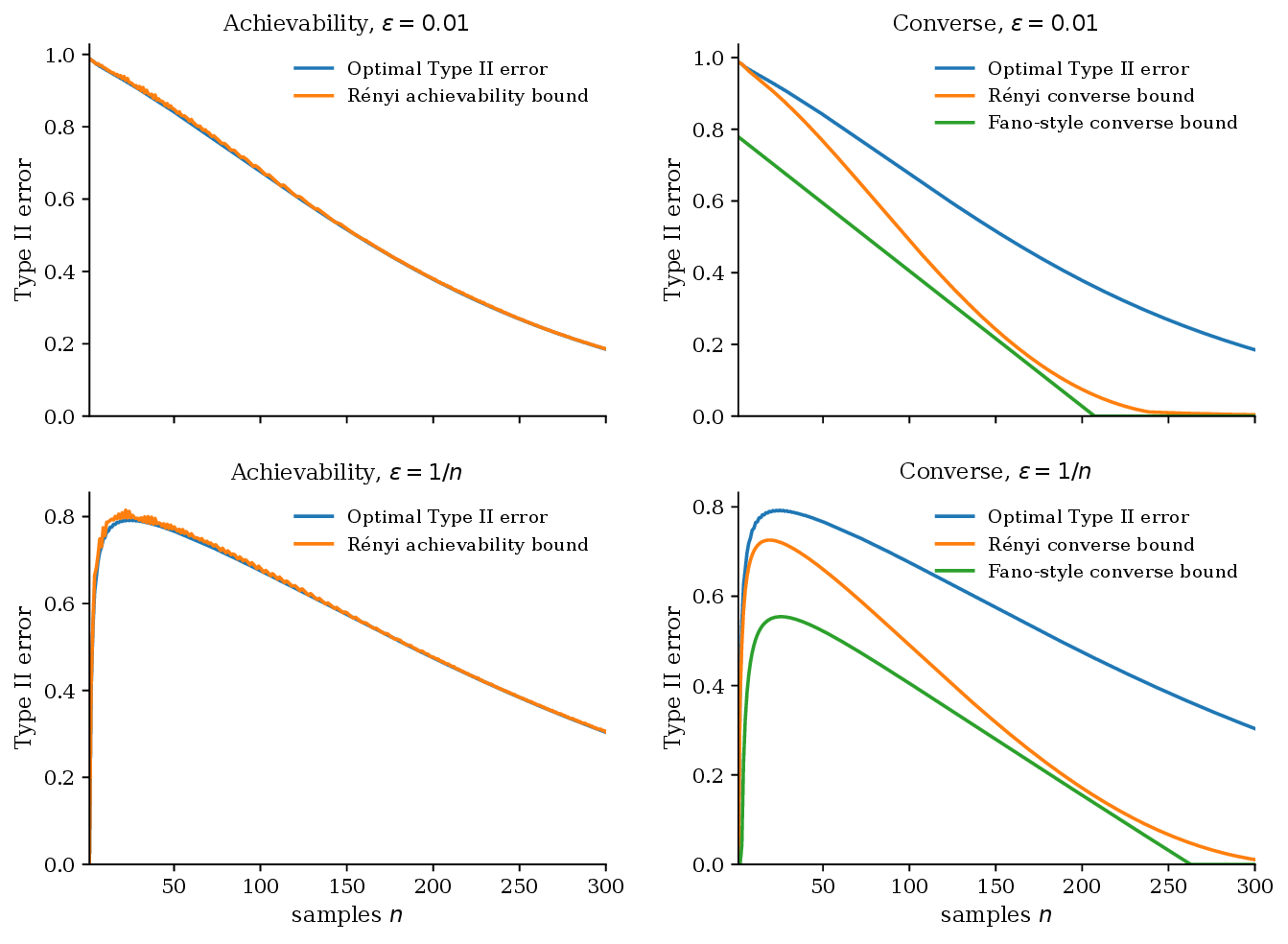}
\caption{Finite sample Type II error and bounds for the nonordered affine ternary classes under fixed and subexponential Type I constraints. The top row uses $\varepsilon_n=0.01$ and the bottom row uses $\varepsilon_n=1/n$. The left column shows the calibrated R\'enyi achievability bound and the right column shows the R\'enyi and the Fano-style converse bounds.}
\label{fig:fixed-subexponential-bounds}
\end{figure}

For achievability, we use the calibrated projected threshold test of \Cref{prop:projected-tightening}. The left column of \Cref{fig:fixed-subexponential-bounds} compares this bound with the optimal Type II error. The top panel uses the fixed constraint $\varepsilon_n=0.01$, while the bottom panel uses the subexponential constraint $\varepsilon_n=1/n$.

For the converse, we use \Cref{thm:finite-blocklength-composite-renyi-converse} together with a Fano-style converse bound. The right column of \Cref{fig:fixed-subexponential-bounds} compares these bounds with the numerical Type II error under each Type I constraint. In both cases, the R\'enyi converse is tighter and remains positive over a larger range of sample sizes. Both converse bounds eventually decrease to zero because $\log(1/\varepsilon_n)=o(n)$.

%##########################################
\section{Conclusion}

We derived finite sample R\'enyi achievability and converse bounds for composite binary hypothesis testing under a uniform Type I constraint ($\varepsilon$). The bounds identify the phase transition under $\varepsilon=e^{-nr}$ and, for compact convex classes with full support on a finite alphabet, give the exact exponents on both sides of the transition, including the strong converse exponent above the phase transition threshold. A central ingredient of the achievability result is the joint R\'enyi projection, whose likelihood ratio provides uniform control over both classes without requiring the projected pair to be least favourable at finite sample size. The same R\'enyi structure recovers the fixed Type I composite Chernoff--Stein exponent through the order zero limit and determines the polynomial dependence on sample size at the selected order. Finally, we identified additional stochastic ordering conditions under which the projected pair is also least favourable, giving an exact finite sample reduction for separated one parameter natural exponential families.

Several questions remain open. For instance, the calibrated achievability bound currently requires numerical optimisation. It remains to be seen whether the tightened bound admits an analytical characterisation and, consequently, a closed form expression. Our polynomial refinement provides a sharper description of its finite sample behaviour by determining the power of $n$ and the logarithmic threshold correction, but does not identify the leading multiplicative constant. On the converse side, our bound applies the finite sample R\'enyi converse to individual pairs $(P^{\otimes n},Q^{\otimes n})$. One may instead consider Bayesian mixtures such as $\int P^{\otimes n}d\Pi_0(P)$ and $\int Q^{\otimes n}d\Pi_1(Q)$, each of which gives a valid lower bound on the composite minimax error. Whether optimising the R\'enyi converse over such mixture pairs yields tighter finite sample bounds remains open. Since the leading exponential rate is already exact under the finite alphabet assumptions, any improvement must occur at finer finite sample scales.

%#########################################
\appendices

%###################################################
\section{Bounds}
\label[appendix]{app:finite-blocklength-bounds}

%###################################################
\subsection{Proof of \Cref{thm:finite-blocklength-composite-renyi-converse}}
\label[appendix]{app:finite-blocklength-composite-renyi-converse}

\begin{IEEEproof}
Fix an admissible test $\varphi_n$, $P\in\mathcal C_0$, $Q\in\mathcal C_1$, and $\lambda>1$. Suppose first that $D_\lambda(Q\|P)<+\infty$. Then $Q\ll P$, and H\"older's inequality, the Type I constraint, and additivity of the R\'enyi divergence give
\begin{equation*}
\begin{aligned}
\mathbb E_{Q^{\otimes n}}[\varphi_n]&=\mathbb E_{P^{\otimes n}}\left[\frac{dQ^{\otimes n}}{dP^{\otimes n}}\varphi_n \right] \leq \exp\left\{-\frac{\lambda-1}{\lambda} \left[\log\frac{1}{\varepsilon}-nD_\lambda(Q\|P) \right]\right\},
\end{aligned}
\end{equation*}
where $\varphi_n^{\lambda/(\lambda-1)}\leq\varphi_n$. Combining this with $\mathbb E_{Q^{\otimes n}}[\varphi_n]\leq1$ gives
\begin{equation*}
\mathbb E_{Q^{\otimes n}}[1-\varphi_n] \geq 1-\exp\left\{ -\frac{\lambda-1}{\lambda} \left[\log\frac{1}{\varepsilon} -nD_\lambda(Q\|P) \right]_+ \right\}.
\end{equation*}
For $D_\lambda(Q\|P)=+\infty$, the bound is trivial. If $D_\lambda(\mathcal C_1\|\mathcal C_0)=+\infty$, the resulting class bound is trivial. Otherwise, choose $(Q_k,P_k)\in\mathcal C_1\times\mathcal C_0$ such that $D_\lambda(Q_k\|P_k)\longrightarrow D_\lambda(\mathcal C_1\|\mathcal C_0).$ Since the preceding inequality holds for every pair, letting $k\to\infty$ gives
\begin{equation*}
\sup_{Q\in\mathcal C_1} \mathbb E_{Q^{\otimes n}}[1-\varphi_n] \geq 1-\exp\left\{-\frac{\lambda-1}{\lambda}\left[\log\frac{1}{\varepsilon} -nD_\lambda(\mathcal C_1\|\mathcal C_0) \right]_+\right\}.
\end{equation*}
Since this holds for every $\lambda>1$, taking the supremum over $\lambda>1$ and then the infimum over all admissible tests yields \eqref{eq:finite-blocklength-composite-renyi-converse}.
\end{IEEEproof}

%###################################################
\subsection{Proof of \Cref{lem:uniform-score-achievability}}
\label[appendix]{app:uniform-score-achievability}

\begin{IEEEproof}
The moment assumptions imply $ P(h=+\infty)=0 \quad \forall P\in\mathcal C_0,$ and $ Q(h=-\infty)=0 \quad \forall Q\in\mathcal C_1,$ because the corresponding exponential would otherwise have infinite expectation. Let $S_n(x^n) := \sum_{i=1}^n h(x_i),$ with $S_n=0$ on sequences containing both $+\infty$ and $-\infty$. This set is null under every relevant product law, so the convention
does not affect either error probability. Independence gives
\begin{equation*}
\sup_{P\in\mathcal C_0} \mathbb E_{P^{\otimes n}} [e^{\lambda S_n}] \leq e^{n(\lambda-1)\xi_0}, \quad {\rm and} \quad \sup_{Q\in\mathcal C_1} \mathbb E_{Q^{\otimes n}} [e^{(\lambda-1)S_n}] \leq e^{n(\lambda-1)\xi_1}.
\end{equation*}
Markov's inequality gives
\begin{equation*}
\alpha_n(\psi_{n,\tau};\mathcal C_0) \leq e^{-\lambda\tau+n(\lambda-1)\xi_0}.
\end{equation*}
Since $\lambda-1<0$, $S_n<\tau$
is equivalent to $e^{(\lambda-1)S_n}>e^{(\lambda-1)\tau}.$ Therefore
\begin{equation*}
\beta_n(\psi_{n,\tau};\mathcal C_1)=
\sup_{Q\in\mathcal C_1}Q^{\otimes n}(S_n<\tau) \leq
e^{(1-\lambda)\tau+n(\lambda-1)\xi_1}.
\end{equation*}
Finally, the Type I bound is at most $\varepsilon$ whenever
\begin{equation*}
\tau \geq \frac{ \log(1/\varepsilon)-n(1-\lambda)\xi_0 }{\lambda}.
\end{equation*}
\end{IEEEproof}

%###################################################
\subsection{\Cref{lem:hellinger-continuity}}
\label[appendix]{app:hellinger-properties}
\begin{lemma}
\label{lem:hellinger-continuity}
Let $0<\alpha<1$. For probability densities
$p,\widetilde p,q,\widetilde q$ with respect to $\mu$,
\begin{equation*}
\left| \int q^\alpha p^{1-\alpha}\,d\mu -\int \widetilde q^\alpha\widetilde p^{1-\alpha}\,d\mu \right| \leq \|q-\widetilde q\|_1^\alpha + \|p-\widetilde p\|_1^{1-\alpha}.
\end{equation*}
Hence $H_\alpha$ is continuous in the product $L^1(\mu)$ norm and weakly upper semicontinuous, while $(Q,P)\mapsto D_\alpha(Q\|P)$ is weakly lower semicontinuous.
\end{lemma}
\begin{IEEEproof}
Insert and subtract $\widetilde q^\alpha p^{1-\alpha}$. Using $|u^\alpha-v^\alpha|\leq|u-v|^\alpha$ and H\"older's inequality,
\begin{equation*}
\begin{aligned}
\left|\int q^\alpha p^{1-\alpha}\,d\mu-\int \widetilde q^\alpha\widetilde p^{1-\alpha}\,d\mu
\right|&\leq \int |q-\widetilde q|^\alpha p^{1-\alpha}\,d\mu+ \int \widetilde q^\alpha
|p-\widetilde p|^{1-\alpha}\,d\mu \leq
\|q-\widetilde q\|_1^\alpha + \|p-\widetilde p\|_1^{1-\alpha}.
\end{aligned}
\end{equation*}
Thus $H_\alpha$ is norm continuous. Its superlevel sets are convex by joint concavity and norm closed by the displayed bound, hence weakly
closed. Therefore $H_\alpha$ is weakly upper semicontinuous. Since $D_\alpha=-\frac{1}{1-\alpha}\log H_\alpha,$ with $D_\alpha=+\infty$ when $H_\alpha=0$, the divergence is weakly lower semicontinuous.
\end{IEEEproof}

%###################################################
\subsection{\Cref{lem:complete-feasible-directional-derivative}}
\label[appendix]{app:feasible-directional-derivative}

The proof of \Cref{thm:projected-finite-blocklength-achievability} uses the following lemma, which allows the maximising function to vanish on sets of positive measure.

\begin{lemma}
\label{lem:complete-feasible-directional-derivative}
Let $0<\alpha<1$, and let $w,x,y$ be nonnegative measurable functions that are finite $\mu$ almost everywhere. For $y_t=(1-t)y+tx$, suppose that
\begin{equation*}
0<\int wy^\alpha\,d\mu<+\infty, \qquad \int wy_t^\alpha\,d\mu \leq \int wy^\alpha\,d\mu \quad \forall t\in[0,1].
\end{equation*}
Then
\begin{equation*}
\mu\{w>0,\ y=0,\ x>0\}=0
\end{equation*}
and
\begin{equation*}
\int_{\{y>0\}}wy^{\alpha-1}x\,d\mu \leq \int wy^\alpha\,d\mu. 
\end{equation*}
Moreover,
\begin{equation}
\lim_{t\downarrow0} \frac{\int wy_t^\alpha\,d\mu-\int wy^\alpha\,d\mu}{t}=\alpha
\left[\int_{\{y>0\}}wy^{\alpha-1}x\,d\mu-\int wy^\alpha\,d\mu\right] \leq 0.
\label{eq:feasible-directional-derivative}
\end{equation}
\end{lemma}

\begin{IEEEproof}
Set $I=\int wy^\alpha\,d\mu, \quad S=\{y>0\}, \quad A=\{w>0,\ y=0\}.$ For $t>0$,
\begin{equation*}
I\geq \int wy_t^\alpha\,d\mu \geq (1-t)^\alpha I + t^\alpha\int_Awx^\alpha\,d\mu,
\end{equation*}
and hence
\begin{equation*}
0 \leq \int_Awx^\alpha\,d\mu \leq I\frac{1-(1-t)^\alpha}{t^\alpha} \leq It^{1-\alpha}.
\end{equation*}
Letting $t\downarrow0$ gives the support assertion. For $r\geq0$, let
\begin{equation*}
g_t(r)=\frac{(1-t+tr)^\alpha-1}{t}.
\end{equation*}
Then
\begin{equation*}
\int_Swy^\alpha g_t(x/y)\,d\mu \leq 0, \quad g_t(r)\geq-1,\quad g_t(r)\to\alpha(r-1).
\end{equation*}
Fatou's lemma gives
\begin{equation*}
(1-\alpha)I + \alpha\int_Swy^{\alpha-1}x\,d\mu \leq I,
\end{equation*}
which proves the integrability bound. Concavity gives $g_t(r)\leq\alpha(r-1)$, so $|g_t(r)|\leq1+\alpha r$. Dominated convergence then yields \eqref{eq:feasible-directional-derivative}.
\end{IEEEproof}

%###################################################
\subsection{Proof of \Cref{thm:projected-finite-blocklength-achievability}}
\label[appendix]{app:projected-finite-blocklength-achievability}

\begin{IEEEproof}
By \Cref{lem:hellinger-continuity}, the Hellinger integral~\Cref{eq:hellinger-integral} attains its maximum on the weakly compact product of the two density classes. Since $\lambda-1<0$, every maximising pair is a joint R\'enyi projection. Write
\begin{equation*}
p_\lambda^\star=\frac{dP_\lambda^\star}{d\mu},
\qquad q_\lambda^\star=\frac{dQ_\lambda^\star}{d\mu},
\end{equation*}
and choose measurable versions
\begin{equation*}
a_\lambda^\star=\frac{dP_\lambda^\star}
{d(P_\lambda^\star+Q_\lambda^\star)},
\qquad b_\lambda^\star =\frac{dQ_\lambda^\star}
{d(P_\lambda^\star+Q_\lambda^\star)}.
\end{equation*}
Define
\begin{equation*}
h_\lambda^\star(x)=
\begin{cases}
\log(b_\lambda^\star(x)/a_\lambda^\star(x)), & a_\lambda^\star(x)>0,\ b_\lambda^\star(x)>0,\\
-\infty, & a_\lambda^\star(x)>0,\ b_\lambda^\star(x)=0,\\
+\infty, & a_\lambda^\star(x)=0,\ b_\lambda^\star(x)>0,\\
0, & a_\lambda^\star(x)=b_\lambda^\star(x)=0.
\end{cases}
\end{equation*}
Fix $P\in\mathcal C_0$ and $Q\in\mathcal C_1$, with densities $p$ and $q$. Optimality of the projected pair and convexity give, for every $t\in[0,1]$,
\begin{equation*}
\begin{aligned}
\int ((1-t)q_\lambda^\star+tq)^\lambda (p_\lambda^\star)^{1-\lambda}\,d\mu &\leq H_\lambda(Q_\lambda^\star,P_\lambda^\star), \qquad \int (q_\lambda^\star)^\lambda ((1-t)p_\lambda^\star+tp)^{1-\lambda}\,d\mu &\leq H_\lambda(Q_\lambda^\star,P_\lambda^\star).
\end{aligned}
\end{equation*}
Since $H_\lambda(Q_\lambda^\star,P_\lambda^\star)\in(0,1]$, \Cref{lem:complete-feasible-directional-derivative} yields
\begin{equation*}
\begin{aligned}
Q\{p_\lambda^\star>0,\ q_\lambda^\star=0\}&=0,
& P\{p_\lambda^\star=0,\ q_\lambda^\star>0\}&=0,\\
\int_{\{p_\lambda^\star q_\lambda^\star>0\}} q(q_\lambda^\star)^{\lambda-1} (p_\lambda^\star)^{1-\lambda}\,d\mu &\leq H_\lambda(Q_\lambda^\star,P_\lambda^\star),
&\int_{\{p_\lambda^\star q_\lambda^\star>0\}}
p(q_\lambda^\star)^\lambda (p_\lambda^\star)^{-\lambda}\,d\mu &\leq H_\lambda(Q_\lambda^\star,P_\lambda^\star).
\end{aligned}
\end{equation*}
The common zero set of $p_\lambda^\star$ and $q_\lambda^\star$ is null under every $R\in\mathcal C_0\cup\mathcal C_1$ by assumption. Together with the preceding support relations, this gives
\begin{equation*}
\mathbb E_P[e^{\lambda h_\lambda^\star}] \leq
H_\lambda(Q_\lambda^\star,P_\lambda^\star), \qquad \mathbb E_Q[e^{(\lambda-1)h_\lambda^\star}] \leq H_\lambda(Q_\lambda^\star,P_\lambda^\star).
\end{equation*}
Therefore
\begin{equation*}
\sup_{P\in\mathcal C_0} \mathbb E_P[e^{\lambda h_\lambda^\star}] \leq e^{(\lambda-1)D_\lambda(\mathcal C_1\|\mathcal C_0)}, \qquad \sup_{Q\in\mathcal C_1} \mathbb E_Q[e^{(\lambda-1)h_\lambda^\star}]\leq e^{(\lambda-1)D_\lambda(\mathcal C_1\|\mathcal C_0)}.
\end{equation*}
Apply \Cref{lem:uniform-score-achievability} with $h=h_\lambda^\star,$ $\xi_0=\xi_1=D_\lambda(\mathcal C_1\|\mathcal C_0),$ and
\begin{equation*}
\tau=\frac{\log(1/\varepsilon)-n(1-\lambda)D_\lambda(\mathcal C_1\|\mathcal C_0)}{\lambda}.
\end{equation*}
The resulting test is admissible and satisfies
\begin{equation*}
\beta_n(\psi_{n,\tau};\mathcal C_1)\leq\exp\left[-\frac{1-\lambda}{\lambda}\left(nD_\lambda(\mathcal C_1\|\mathcal C_0)-\log\frac{1}{\varepsilon}\right)\right].
\end{equation*}
The constant randomised test $\varphi_n\equiv\varepsilon$ has Type II error $1-\varepsilon$. Taking the smaller of the two bounds yields \eqref{eq:projected-finite-blocklength-achievability}.
\end{IEEEproof}
%###################################################
\subsection{Proof of \Cref{cor:threshold-rate}}
\label[appendix]{app:threshold-rate}

\begin{IEEEproof} 
Set $\varepsilon=e^{-nr}$. If $0<r<D(\mathcal C_1\|\mathcal C_0)$, then $D_{\lambda_k}(\mathcal C_1\|\mathcal C_0)\to D(\mathcal C_1\|\mathcal C_0)$, so for some fixed $k$, $D_{\lambda_k}(\mathcal C_1\|\mathcal C_0)>r$. By \Cref{thm:projected-finite-blocklength-achievability},
\begin{equation*} 
\beta_n^\star(e^{-nr};\mathcal C_0,\mathcal C_1) \leq \exp\left\{ -n\frac{1-\lambda_k}{\lambda_k} \left[D_{\lambda_k}(\mathcal C_1\|\mathcal C_0)-r\right] \right\}.
\end{equation*} 
The exponent coefficient is strictly positive, hence $\beta_n^\star(e^{-nr};\mathcal C_0,\mathcal C_1)\to0$. If $r>D(\mathcal C_1\|\mathcal C_0)$, the limit from above gives some $\lambda>1$ such that $D_\lambda(\mathcal C_1\|\mathcal C_0)<r$. By \Cref{thm:finite-blocklength-composite-renyi-converse},
\begin{equation*}
\beta_n^\star(e^{-nr};\mathcal C_0,\mathcal C_1) \geq 1-\exp\left\{-n\frac{\lambda-1}{\lambda} \left[r-D_\lambda(\mathcal C_1\|\mathcal C_0)\right]\right\}.
\end{equation*} 
Again the exponent coefficient is strictly positive, hence $\beta_n^\star(e^{-nr};\mathcal C_0,\mathcal C_1)\to1$.
\end{IEEEproof}

%###################################################
\subsection{Proof of \Cref{lem:composite-renyi-order-one}}
\label[appendix]{app:composite-renyi-order-one}

\begin{IEEEproof}
Monotonicity gives $D_\lambda(\mathcal C_1\|\mathcal C_0) \geq D(\mathcal C_1\|\mathcal C_0) \qquad \forall\lambda>1.$ Fix $\eta>0$ and choose $(Q_\eta,P_\eta,\delta_\eta)$ as in the assumptions. Since $D_{1+\delta_\eta}(Q_\eta\|P_\eta)<+\infty,$ the fixed pair R\'enyi to KL limit gives, for all $\lambda>1$ sufficiently close to one,
\begin{equation*}
D_\lambda(\mathcal C_1\|\mathcal C_0) \leq D_\lambda(Q_\eta\|P_\eta) \leq D(Q_\eta\|P_\eta)+\eta \leq D(\mathcal C_1\|\mathcal C_0)+2\eta.
\end{equation*}
Letting $\lambda\downarrow 1$ and then $\eta\downarrow0$ proves the result.
\end{IEEEproof}

%###################################################
\subsection{Corollary~\ref{cor:composite-renyi-order-one-below}}
\label[appendix]{app:composite-renyi-order-one-below}

\begin{corollary}
\label{cor:composite-renyi-order-one-below}
Suppose that all laws in $\mathcal C_0$ and $\mathcal C_1$ admit densities with respect to a common $\sigma$-finite measure $\mu$, and that, when identified with their densities with respect to $\mu$, $\mathcal C_0$ and $\mathcal C_1$ are weakly compact in $L^1(\mu)$. Then $ \lim_{\lambda\uparrow1} D_\lambda(\mathcal C_1\|\mathcal C_0) = D(\mathcal C_1\|\mathcal C_0).$
\end{corollary}

\begin{IEEEproof}
Monotonicity in the R\'enyi order gives $D_\lambda(\mathcal C_1\|\mathcal C_0) \uparrow L \leq D(\mathcal C_1\|\mathcal C_0)$ as $\lambda\uparrow1$ for some $L\in[0,+\infty]$. Suppose that $L<D(\mathcal C_1\|\mathcal C_0)$ and choose $c$ strictly between them. By \Cref{lem:hellinger-continuity}, the sets $\mathcal A_\lambda=\{(Q,P)\in\mathcal C_1\times\mathcal C_0 \allowbreak \mid D_\lambda(Q\|P)\leq c\}$. are weakly closed. They are nonempty, weakly compact, and nested as $\lambda\uparrow1$. Their intersection is therefore nonempty. Choose $(Q^\star,P^\star)$ in the intersection. Then $D_\lambda(Q^\star\|P^\star)\leq c \quad \forall\lambda<1.$ Letting $\lambda\uparrow1$ gives $D(Q^\star\|P^\star)\leq c,$ contradicting
$c<D(\mathcal C_1\|\mathcal C_0)$. Hence, $L=D(\mathcal C_1\|\mathcal C_0).$
\end{IEEEproof}

%###################################################
\section{Exact error exponents on finite alphabets}
\label[appendix]{app:finite-alphabet-consequences}

%###################################################
\subsection{\Cref{lem:simple-pair-exponent}}
\label[appendix]{app:simple-pair-exponent}

\begin{lemma}
\label{lem:simple-pair-exponent} 
Let $\mathcal X$ be a finite alphabet with $d=|\mathcal X|$, let $P$ and $Q$ have full support, and let $0<r<D(Q\|P)$. Then
\begin{equation*} 
\lim_{n\to\infty} -\frac{1}{n} \log \beta_n^\star(e^{-nr};P,Q) = \max_{0<s<1} \frac{1-s}{s} \left[ D_s(Q\|P)-r \right].
\end{equation*}
\end{lemma}

\begin{IEEEproof}
Set $\psi(s) = \log H_s(Q,P).$ For $0<s<1$, the likelihood ratio test with threshold $n[\psi(s)+r]/s$ has Type I error at most $e^{-nr}$ and Type II error at most
\begin{equation*} 
\exp\left\{ -n\frac{1-s}{s} \left[ D_s(Q\|P)-r \right] \right\}.
\end{equation*}
Optimising over $s$ proves achievability. For the converse, define
\begin{equation*} 
R_s(x) = \frac{ Q(x)^sP(x)^{1-s} }{ H_s(Q,P) }.
\end{equation*}
Direct differentiation gives
\begin{equation*} 
\frac{d}{ds} D(R_s\|P) = s\, \operatorname{Var}_{R_s} \left( \log\frac{Q}{P} \right)
\end{equation*}
and
\begin{equation*} 
\frac{d}{ds} \left\{ \frac{1-s}{s} \left[ D_s(Q\|P)-r \right] \right\} = \frac{ r-D(R_s\|P) }{s^2}.
\end{equation*}
Since $D(Q\|P)>r>0$, there is a unique $s_r\in(0,1)$ satisfying $D(R_{s_r}\|P)=r.$
It is the maximising order and
\begin{equation*} 
\max_{0<s<1} \frac{1-s}{s} \left[ D_s(Q\|P)-r \right] = D(R_{s_r}\|Q).
\end{equation*}
Fix $0<t<s_r$, put $V=R_t$, and choose $n$ types $V_n\to V$. After symmetrisation, any admissible test is constant, say $c_n$, on the type class $\mathcal T_{V_n}$. Standard type bounds and the Type I constraint give
$c_n \leq (n+1)^d e^{-n[r-D(V_n\|P)]} =: \nu_n,$ as $\nu_n\longrightarrow0.$ Hence
\begin{equation*} 
\mathbb E_{Q^{\otimes n}}[1-\varphi_n] \geq (1-\nu_n) (n+1)^{-d} e^{-nD(V_n\|Q)}.
\end{equation*} 
Taking the infimum over admissible tests, then $n\to\infty$ and $t\uparrow s_r$, gives the matching converse exponent.
\end{IEEEproof}

%###################################################
\subsection{Proof of \Cref{thm:exact-composite-exponent}}
\label[appendix]{app:exact-composite-exponent}

\begin{IEEEproof}
For $\rho\geq0$, set $ \lambda_\rho=1/(1+\rho)$ and
\begin{equation*}
G_r(\rho;P,Q)=
\begin{cases}
\rho\left[ D_{\lambda_\rho}(Q\|P)-r \right], & \rho>0, \\ 0, & \rho=0.
\end{cases}
\end{equation*}
For $0<\lambda<1$, define
\begin{equation*} 
R_{\lambda;P,Q}(x) = \frac{ Q(x)^\lambda P(x)^{1-\lambda} }{ H_\lambda(Q,P) }.
\end{equation*}
The identity $ (1-\lambda)D(V\|P) + \lambda D(V\|Q) = D(V\|R_{\lambda;P,Q}) - \log H_\lambda(Q,P) $ gives
\begin{equation} 
G_r(\rho;P,Q) = \min_{V\in\Delta_d} \left\{ D(V\|Q) + \rho \left[ D(V\|P)-r \right] \right\}.
\label{eq:proof-rho-representation}
\end{equation} 
The representation shows that $G_r$ is concave in $\rho$, while joint convexity of $D_{\lambda_\rho}$ gives joint convexity in $(P,Q)$ for $\rho>0$~\cite[Th.~11]{van2014renyi}. Uniform full support and compactness give
\begin{equation*} 
K = \max_{\substack{P\in\mathcal C_0\\Q\in\mathcal C_1}} \max \left\{ D(P\|Q), D(Q\|P) \right\} <+\infty.
\end{equation*}
Since $0\leq D_\lambda(Q\|P)\leq D(Q\|P)$, $G_r$ is jointly continuous on every compact $\rho$ interval. Choosing $V=P$ in \eqref{eq:proof-rho-representation} gives $G_r(\rho;P,Q) \leq D(P\|Q)-\rho r \leq K-\rho r.$ Choose $\overline\rho>K/r$. All relevant maxima may then be restricted to $[0,\overline\rho]$. Sion's minimax theorem gives
\begin{equation} 
\max_{0\leq\rho\leq\overline\rho} \min_{\substack{P\in\mathcal C_0\\Q\in\mathcal C_1}} G_r(\rho;P,Q) = \min_{\substack{P\in\mathcal C_0\\Q\in\mathcal C_1}} \max_{0\leq\rho\leq\overline\rho} G_r(\rho;P,Q).
\label{eq:proof-minimax-interchange}
\end{equation}

Let $\rho_r^\star$ maximise the left side and let $(P_r^\star,Q_r^\star)$ minimise the right side. If $v_r$ denotes the common value, then
\begin{equation*} 
v_r = \min_{P,Q} G_r(\rho_r^\star;P,Q) \leq G_r(\rho_r^\star;P_r^\star,Q_r^\star) \leq \max_\rho G_r(\rho;P_r^\star,Q_r^\star) = v_r.
\end{equation*}
Hence equality holds throughout and the resulting triple is a saddle point. Because $r<D(\mathcal C_1\|\mathcal C_0)$, Corollary~\ref{cor:composite-renyi-order-one-below} gives some $\lambda<1$ such that $D_\lambda(\mathcal C_1\|\mathcal C_0)>r.$ Thus $v_r>0$. Since $G_r(0;P,Q)=0$ and $G_r(\overline\rho;P,Q)<0$ for every pair,
$ 0<\rho_r^\star<\overline\rho.$ The bound $G_r(\rho;P,Q)\leq K-\rho r$ extends the left saddle inequality to all $\rho\geq0$.

Set $\lambda_r^\star =1/(1+\rho_r^\star).$ Under $\rho=(1-\lambda)/\lambda$, the saddle inequalities become
\begin{equation}
\begin{aligned} 
\frac{1-\lambda}{\lambda} \left[ D_\lambda(Q_r^\star\|P_r^\star)-r \right] &\leq \frac{1-\lambda_r^\star}{\lambda_r^\star} \left[ D_{\lambda_r^\star}(Q_r^\star\|P_r^\star)-r \right]  \leq \frac{1-\lambda_r^\star}{\lambda_r^\star} \left[ D_{\lambda_r^\star}(Q\|P)-r \right]
\end{aligned}
\label{eq:proof-exact-exponent-saddle}
\end{equation} 
for every $0<\lambda<1$, $P\in\mathcal C_0$, and $Q\in\mathcal C_1$. The right inequality shows that $(Q_r^\star,P_r^\star)$ is projected at order $\lambda_r^\star$. The change of variables in \eqref{eq:proof-minimax-interchange} also gives the two variational expressions in \eqref{eq:exact-composite-exponent}. The projected bound at $\lambda_r^\star$ gives
\begin{equation*} 
\liminf_{n\to\infty} -\frac{1}{n} \log \beta_n^\star(e^{-nr};\mathcal C_0,\mathcal C_1) \geq G_r(\rho_r^\star;P_r^\star,Q_r^\star).
\end{equation*}
Conversely,
\begin{equation*} 
\beta_n^\star(e^{-nr};\mathcal C_0,\mathcal C_1) \geq \beta_n^\star(e^{-nr};P_r^\star,Q_r^\star).
\end{equation*}
Since $v_r>0$, $D_{\lambda_r^\star}(Q_r^\star\|P_r^\star)>r.$ Monotonicity in the R\'enyi order therefore gives $D(Q_r^\star\|P_r^\star)>r.$ \Cref{lem:simple-pair-exponent} and the left saddle inequality therefore give
\begin{equation*} 
\lim_{n\to\infty} -\frac{1}{n} \log \beta_n^\star(e^{-nr};P_r^\star,Q_r^\star) = G_r(\rho_r^\star;P_r^\star,Q_r^\star).
\end{equation*}
This proves \eqref{eq:exact-composite-exponent} and the asserted
attainment. For later use, define
\begin{equation} 
R_r^\star(x) = \frac{ Q_r^\star(x)^{\lambda_r^\star} P_r^\star(x)^{1-\lambda_r^\star} }{ H_{\lambda_r^\star}(Q_r^\star,P_r^\star) }.
\label{eq:proof-exact-exponent-distribution}
\end{equation} 
More generally, let $R_\rho$ denote the same tilted distribution with order $\lambda_\rho$. Direct differentiation gives
\begin{equation*}
\frac{d}{d\rho} G_r(\rho;P_r^\star,Q_r^\star) = D(R_\rho\|P_r^\star)-r.
\end{equation*}
The derivative vanishes at the interior maximiser, so
\begin{equation} 
D(R_r^\star\|P_r^\star) = r.
\label{eq:proof-exact-exponent-rate-match}
\end{equation}
Substitution into \eqref{eq:proof-rho-representation} gives
\begin{equation} 
D(R_r^\star\|Q_r^\star) = \frac{1-\lambda_r^\star}{\lambda_r^\star} \left[ D_{\lambda_r^\star}(Q_r^\star\|P_r^\star)-r \right].
\label{eq:proof-exact-exponent-divergence}
\end{equation}
\end{IEEEproof}

%###################################################
\subsection{Proof of \Cref{cor:hoeffding-tangent}}
\label[appendix]{app:hoeffding-tangent}
 
\begin{IEEEproof} For the proof, define $ E(r) := \max_{0<\lambda<1} \frac{1-\lambda}{\lambda} \left[ D_\lambda(\mathcal C_1\|\mathcal C_0)-r \right]$ and write $ d(\lambda) = D_\lambda(\mathcal C_1\|\mathcal C_0).$ Set \begin{equation*} \rho = \frac{1-\lambda}{\lambda}, \qquad \lambda = \frac{1}{1+\rho}, \end{equation*} and define \begin{equation*} \Gamma(\rho) = \rho\, d\left( \frac{1}{1+\rho} \right). \end{equation*} Hence \begin{equation} E(r) = \sup_{\rho>0} \left\{ \Gamma(\rho)-\rho r \right\}. \label{eq:appendix-hoeffding-rho} \end{equation}
Fix $0<r<D(\mathcal C_1\|\mathcal C_0).$ The order one limit from below gives $\Gamma(\rho)\to0$ as $\rho\downarrow0$, while it also gives a value $\rho_0>0$ such that $\Gamma(\rho_0)-\rho_0 r>0.$ These inequalities remain uniform for all rates in a sufficiently small neighbourhood of $r$. Hence all maximisers are bounded away from zero throughout that neighbourhood.
 
For the other endpoint, Jensen's inequality gives, for every $P\in\mathcal C_0$, $Q\in\mathcal C_1$, $(1-\lambda)/(\lambda) D_\lambda(Q\|P) \leq D(P\|Q).$ Where $0<\lambda<1$. Compactness and common full support imply \begin{equation*} K = \sup_{\substack{P\in\mathcal C_0\\Q\in\mathcal C_1}} D(P\|Q) <+\infty. \end{equation*} Thus, for rates in a sufficiently small neighbourhood of $r$, $\Gamma(\rho)-\rho r' \leq K-\rho r' \longrightarrow -\infty$ uniformly as $\rho\to\infty$. Therefore all maximisers lie in a common compact interval $0<\rho_-<\rho<\rho_+<+\infty.$ For a fixed pair $(P,Q)$ define $\gamma_{P,Q}(\rho) = \rho D_{1/(1+\rho)}(Q\|P).$ Since \begin{equation*} d(\lambda) = \min_{\substack{P\in\mathcal C_0\\Q\in\mathcal C_1}} D_\lambda(Q\|P), \end{equation*} we have \begin{equation} \Gamma(\rho) = \min_{\substack{P\in\mathcal C_0\\Q\in\mathcal C_1}} \gamma_{P,Q}(\rho). \label{eq:appendix-gamma-min} \end{equation}

Let $\lambda_\rho = 1/(1+\rho)$ and define \begin{equation*} R_{\lambda_\rho;P,Q}(x) = \frac{ Q(x)^{\lambda_\rho} P(x)^{1-\lambda_\rho} }{ H_{\lambda_\rho}(Q,P) }. \end{equation*} Direct differentiation gives \begin{equation} \gamma_{P,Q}''(\rho) = -\lambda_\rho^3 \operatorname{Var}_{R_{\lambda_\rho;P,Q}} \left( \log\frac{Q}{P} \right). \label{eq:appendix-gamma-curvature} \end{equation}
 
Since $D(\mathcal C_1\|\mathcal C_0)>r>0$, no distribution in $\mathcal C_0$ coincides with one in $\mathcal C_1$. The variance in \eqref{eq:appendix-gamma-curvature} is therefore strictly positive for every pair. Continuity and compactness give $c>0$ such that $\gamma_{P,Q}''(\rho) \leq -c$ uniformly over all pairs and $\rho\in[\rho_-,\rho_+]$. Thus every $\gamma_{P,Q}$ is strongly concave with the same constant. Consequently, \begin{equation*} \Gamma(\rho) + \frac{c}{2}\rho^2 = \min_{\substack{P\in\mathcal C_0\\Q\in\mathcal C_1}} \left[ \gamma_{P,Q}(\rho) + \frac{c}{2}\rho^2 \right] \end{equation*} is concave. Hence $\Gamma$ is strongly concave and $\rho \longmapsto \Gamma(\rho)-\rho r$ has a unique maximiser $\rho_r^\star$. Therefore
\begin{equation*}
\lambda_r^\star= \frac{1}{1+\rho_r^\star}
\end{equation*}
is unique. Danskin's theorem applied to \eqref{eq:appendix-hoeffding-rho} gives 
\begin{equation*} 
E'(r) = -\rho_r^\star = -\frac{1-\lambda_r^\star}{\lambda_r^\star}. 
\end{equation*} 
Thus the derivative of the exact Type II error exponent is 
\begin{equation*} 
-\frac{1-\lambda_r^\star}{\lambda_r^\star}. 
\end{equation*}
Finally, $E$ is convex as the supremum of affine functions of $r$. Since it is differentiable throughout the open achievable regime, its derivative is continuous there. Hence the exact Type II error exponent is continuously differentiable for $0<r<D(\mathcal C_1\|\mathcal C_0).$
\end{IEEEproof}

%###################################################
\subsection{Proof of \Cref{thm:exact-composite-strong-converse-exponent}}
\label[appendix]{app:exact-composite-strong-converse-exponent}

\begin{IEEEproof}
Write
$\Gamma_n(r):=1-\beta_n^\star(e^{-nr};\mathcal C_0,\mathcal C_1)$. Then
\begin{equation}
\Gamma_n(r)
=
\sup_{\substack{
0\leq\varphi_n\leq1\\
\sup_{P\in\mathcal C_0}
\mathbb E_{P^{\otimes n}}\varphi_n
\leq e^{-nr}
}}
\inf_{Q\in\mathcal C_1}
\mathbb E_{Q^{\otimes n}}\varphi_n.
\label{eq:app-strong-converse-gamma}
\end{equation}
Let $\mathcal P_n(\mathcal X)$ denote the set of $n$ types on
$\mathcal X$. For $V\in\Delta_d$, define
$d_{\mathcal C_0}(V):=\inf_{P\in\mathcal C_0}D(V\|P)$ and
$a_r(V):=[r-d_{\mathcal C_0}(V)]_+$. For $Q\in\mathcal C_1$, put
$F_n(Q):=\min_{V\in\mathcal P_n(\mathcal X)}
\{D(V\|Q)+a_r(V)\}$ and
$E_n(r):=\sup_{Q\in\mathcal C_1}F_n(Q)$. Standard type bounds give, for every $n$ type $V$ and every full support
law $R$,
\begin{equation}
(n+1)^{-d}e^{-nD(V\|R)}
\leq
R^{\otimes n}(T_V)
\leq
e^{-nD(V\|R)},
\label{eq:app-type-bounds}
\end{equation}
while $|\mathcal P_n(\mathcal X)|\leq(n+1)^d$. For a lower bound on $\Gamma_n(r)$, define
\begin{equation}
\varphi_n(x^n)
=
(n+1)^{-d}
\exp\{-na_r(\widehat P_{x^n})\}.
\label{eq:app-type-test}
\end{equation}
For every $P\in\mathcal C_0$,
\begin{equation*}
\mathbb E_{P^{\otimes n}}\varphi_n
\leq
(n+1)^{-d}
\sum_{V\in\mathcal P_n(\mathcal X)}
\exp\{-n[D(V\|P)+a_r(V)]\}.
\end{equation*}
Since $D(V\|P)+a_r(V)\geq r$ for every $V$ and $P\in\mathcal C_0$, we have
$\sup_{P\in\mathcal C_0}\mathbb E_{P^{\otimes n}}\varphi_n
\leq e^{-nr}$, so the test is admissible. Fix $Q\in\mathcal C_1$ and choose an $n$ type $V_Q$ attaining
$F_n(Q)$. The lower type bound gives
$\mathbb E_{Q^{\otimes n}}\varphi_n
\geq(n+1)^{-2d}e^{-nF_n(Q)}$. Taking the infimum over $Q$ gives
\begin{equation}
\Gamma_n(r)
\geq
(n+1)^{-2d}e^{-nE_n(r)}.
\label{eq:app-gamma-lower}
\end{equation}
For the upper bound, symmetrise any admissible test over coordinate
permutations. Its value on a type class $T_V$ may then be written as
$t_V\in[0,1]$. Compactness of $\mathcal C_0$ gives
$P_V\in\mathcal C_0$ satisfying
$D(V\|P_V)=d_{\mathcal C_0}(V)$. The Type I constraint and the lower
type bound give
$e^{-nr}\geq t_V(n+1)^{-d}e^{-nd_{\mathcal C_0}(V)}$.
Combining this with $t_V\leq1$ yields
\begin{equation}
t_V
\leq
(n+1)^d e^{-na_r(V)}.
\label{eq:app-type-test-upper}
\end{equation}
Therefore, for every $Q\in\mathcal C_1$,
\begin{equation*}
\mathbb E_{Q^{\otimes n}}\varphi_n
\leq
(n+1)^d
\sum_{V\in\mathcal P_n(\mathcal X)}
\exp\{-n[D(V\|Q)+a_r(V)]\}
\leq
(n+1)^{2d}e^{-nF_n(Q)}.
\end{equation*}
Since $F_n$ is continuous on the compact class $\mathcal C_1$, its
supremum is attained. Choosing a maximising $Q$ and then taking the
supremum over admissible tests gives
\begin{equation}
\Gamma_n(r)
\leq
(n+1)^{2d}e^{-nE_n(r)}.
\label{eq:app-gamma-upper}
\end{equation}
Hence
\begin{equation}
(n+1)^{-2d}e^{-nE_n(r)}
\leq
\Gamma_n(r)
\leq
(n+1)^{2d}e^{-nE_n(r)}.
\label{eq:app-type-sandwich}
\end{equation}
Define
$F(Q):=\min_{V\in\Delta_d}\{D(V\|Q)+a_r(V)\}$ and
$E_{\rm type}(r):=\sup_{Q\in\mathcal C_1}F(Q)$.
Uniform full support implies joint continuity of all relevant
divergences on the compact domains. Thus $d_{\mathcal C_0}(V)$ and
$a_r(V)$ are continuous. Since the $n$ types become dense in
$\Delta_d$, uniform continuity gives
$\sup_{Q\in\mathcal C_1}|F_n(Q)-F(Q)|\to0$, and hence
$E_n(r)\to E_{\rm type}(r)$. Using \eqref{eq:app-type-sandwich},
\begin{equation}
\lim_{n\to\infty}
-\frac{1}{n}\log\Gamma_n(r)
=
E_{\rm type}(r).
\label{eq:app-type-exponent}
\end{equation}
It remains to identify $E_{\rm type}(r)$. For fixed
$Q\in\mathcal C_1$, $[x]_+=\sup_{0\leq a\leq1}ax$ and
$r-d_{\mathcal C_0}(V)
=\sup_{P\in\mathcal C_0}[r-D(V\|P)]$. Hence
\begin{equation}
F(Q)
=
\min_{V\in\Delta_d}
\sup_{\substack{0\leq a\leq1\\P\in\mathcal C_0}}
\left\{
D(V\|Q)
+
a[r-D(V\|P)]
\right\}.
\label{eq:app-minmax-start}
\end{equation}
Fix $\delta\in(0,1)$ and let
$\mathcal K_\delta
:=\{(a,S)\mid 0\leq a\leq1-\delta,\ S=aP,\ P\in\mathcal C_0\}$.
Convexity and compactness of $\mathcal C_0$ make
$\mathcal K_\delta$ convex and compact. For $a>0$, put
$\Phi_Q(V,a,S):=D(V\|Q)+ar-aD(V\|S/a)$, with its continuous extension
at $a=0$. For fixed $(a,S)$, $\Phi_Q$ is convex in $V$. For fixed $V$, it is
concave in $(a,S)$ because
$(a,S)\mapsto aD(V\|S/a)$ is the perspective of the convex function
$P\mapsto D(V\|P)$. Sion's minimax theorem gives
\begin{equation}
\min_V
\max_{(a,S)\in\mathcal K_\delta}
\Phi_Q(V,a,S)
=
\max_{(a,S)\in\mathcal K_\delta}
\min_V
\Phi_Q(V,a,S).
\label{eq:app-sion}
\end{equation}

For $a\in(0,1)$, set
$\lambda:=1/(1-a)>1$, so that
$a=(\lambda-1)/\lambda$. A direct variational calculation gives
\begin{equation}
\min_{V\in\Delta_d}
\left\{
D(V\|Q)
+
a[r-D(V\|P)]
\right\}
=
\frac{\lambda-1}{\lambda}
[r-D_\lambda(Q\|P)].
\label{eq:app-gibbs}
\end{equation}
The minimising distribution is proportional to
$Q(x)^\lambda P(x)^{1-\lambda}$. Letting $\delta\downarrow0$ and retaining the $a=0$ branch gives
\begin{equation}
F(Q)
=
\sup_{P\in\mathcal C_0}
\sup_{\lambda>1}
\frac{\lambda-1}{\lambda}
[r-D_\lambda(Q\|P)]_+.
\label{eq:app-FQ-renyi}
\end{equation}
Therefore
\begin{align}
E_{\rm type}(r)=\sup_{\substack{Q\in\mathcal C_1\\P\in\mathcal C_0}}
\sup_{\lambda>1}
\frac{\lambda-1}{\lambda}
[r-D_\lambda(Q\|P)]_+\sup_{\lambda>1}
\frac{\lambda-1}{\lambda}
[r-D_\lambda(\mathcal C_1\|\mathcal C_0)]_+.
\label{eq:app-Etype-renyi}
\end{align}
Combining \eqref{eq:app-type-exponent} and
\eqref{eq:app-Etype-renyi} proves
\eqref{eq:exact-composite-strong-converse-exponent}. The equivalent
asymptotic form \eqref{eq:strong-converse-exact-asymptotic} follows
immediately.
\end{IEEEproof}

%###################################################
\subsection{Proof of \Cref{lem:order-zero-limit}}
\label[appendix]{app:order-zero-limit}

\begin{IEEEproof}
Let $M = \log \Big( \min_{\substack{ R\in\mathcal C_0\cup\mathcal C_1\\ x\in\mathcal X}} R(x) \Big)^{-1}.$ For $P\in\mathcal C_0$ and $Q\in\mathcal C_1$, let $h_{P,Q}(x) = \log Q(x)/P(x).$ Uniform full support gives $|h_{P,Q}|\leq M$, while $\mathbb E_P[h_{P,Q}] = -D(P\|Q)$ and $H_\lambda(Q,P) = \mathbb E_P \left[ e^{\lambda h_{P,Q}} \right].$ Jensen's inequality and Hoeffding's lemma give
\begin{equation*} 
0 \leq \log \mathbb E_P \left[ e^{\lambda h_{P,Q}} \right] - \lambda \mathbb E_P[h_{P,Q}] \leq \frac{M^2}{2}\lambda^2.
\end{equation*}
Since
\begin{equation*} 
\frac{1-\lambda}{\lambda} D_\lambda(Q\|P) = -\frac{1}{\lambda} \log H_\lambda(Q,P),
\end{equation*}
we obtain
\begin{equation*} 
D(P\|Q) - \frac{M^2}{2}\lambda \leq \frac{1-\lambda}{\lambda} D_\lambda(Q\|P) \leq D(P\|Q).
\end{equation*} 
Taking the infimum over $(P,Q)\in\mathcal C_0\times\mathcal C_1$ and letting $\lambda\downarrow0$ proves \eqref{eq:order-zero-limit}.
\end{IEEEproof}

%###################################################
\subsection{Proof of \Cref{thm:fixed-type-one-exponent}}
\label[appendix]{app:fixed-type-one-exponent}

\begin{IEEEproof}
For every $P\in\mathcal C_0$ and $Q\in\mathcal C_1$, pairwise reduction and the Chernoff Stein lemma give
\begin{equation*} 
\limsup_{n\to\infty} -\frac{1}{n} \log \beta_n^\star(\varepsilon;\mathcal C_0,\mathcal C_1) \leq D(P\|Q).
\end{equation*} 
Taking the infimum over the pair gives the converse exponent $D(\mathcal C_0\|\mathcal C_1)$. For achievability, set$ \lambda_n = n^{-1/2}$  and let $M = \log \Big( \min_{\substack{ R\in\mathcal C_0\cup\mathcal C_1\\ x\in\mathcal X}} R(x) \Big)^{-1}.$ \Cref{thm:projected-finite-blocklength-achievability} and \Cref{lem:order-zero-limit} give, for $n\geq2$,
\begin{equation*}
\begin{aligned} 
-\frac{1}{n} \log \beta_n^\star(\varepsilon;\mathcal C_0,\mathcal C_1) &\geq \frac{1-\lambda_n}{\lambda_n} D_{\lambda_n}(\mathcal C_1\|\mathcal C_0) - \frac{1-\lambda_n}{n\lambda_n} \log\frac{1}{\varepsilon} \geq D(\mathcal C_0\|\mathcal C_1) - \frac{M^2}{2\sqrt n} - \frac{1-\lambda_n}{n\lambda_n} \log\frac{1}{\varepsilon}.
\end{aligned}
\end{equation*} The final two terms vanish, proving \eqref{eq:fixed-type-one-exponent}. If $D(\mathcal C_0\|\mathcal C_1)=0$, continuity on the compact product and full support give $P_0=Q_0=R$ for some pair attaining the infimum. Every admissible test has Type II error at least $1-\varepsilon$ under $R^{\otimes n}$, while the constant test $\varphi_n\equiv\varepsilon$ attains equality.
\end{IEEEproof}

%###################################################
\subsection{Proof of \Cref{cor:zero-rate-boundary}}
\label[appendix]{app:zero-rate-boundary}

\begin{IEEEproof}
Let $M = \log \Big(\min_{\substack{ R\in\mathcal C_0\cup\mathcal C_1\\ x\in\mathcal X}} R(x).\Big)^{-1}.$ \Cref{lem:order-zero-limit} gives, for every $0<\lambda<1$,
\begin{equation*} 
\frac{1-\lambda}{\lambda} D_\lambda(\mathcal C_1\|\mathcal C_0) \leq D(\mathcal C_0\|\mathcal C_1),
\end{equation*}
which gives the required upper bound. For sufficiently small $r>0$, take $\lambda=\sqrt r$. The lower estimate in the same lemma gives
\begin{equation*}
\begin{aligned} 
\max_{0<\lambda<1} \frac{1-\lambda}{\lambda} \left[ D_\lambda(\mathcal C_1\|\mathcal C_0)-r \right] &\geq D(\mathcal C_0\|\mathcal C_1) - \frac{M^2}{2}\sqrt r - \frac{1-\sqrt r}{\sqrt r}r =D(\mathcal C_0\|\mathcal C_1) - \left( \frac{M^2}{2}+1 \right) \sqrt r +r.
\end{aligned}
\end{equation*}
Letting $r\downarrow0$ proves the result.
\end{IEEEproof}

%###################################################
\section{Refining the analysis}
\label[appendix]{app:finite-blocklength-exactness}

%###################################################
\subsection{Proof of~\Cref{prop:projected-tightening}}
\label[appendix]{app:projected-tightening}

\begin{IEEEproof}
For the fixed $\lambda$ and projected pair in proposition~\ref{prop:projected-tightening}, define $S^\star(x^n):=\sum_{i=1}^n h_\lambda^\star(x_i).$ Since $\mathcal X$ is finite, $S^\star$ takes finitely many distinct values $v_1<\cdots<v_m.$ For $1\leq j\leq m$ and $0\leq\eta\leq1$, set
\begin{equation*}
\psi_{j,\eta}:=\mathbbm 1\{S^\star>v_j\}+\eta\mathbbm 1\{S^\star=v_j\}.
\end{equation*}
Starting from $\psi_{m,0}=0$, increase $\eta$ at each boundary in the order $v_m,\ldots,v_1$, ending at $\psi_{1,1}=1$. Adjacent segments join because $\psi_{j,1}=\psi_{j-1,0}$. For each boundary, define
\begin{equation*}
A_j(\eta):=\sup_{P\in\mathcal C_0} \mathbb E_{P^{\otimes n}}\left[\psi_{j,\eta}\right].
\end{equation*}
If $0\leq\eta\leq\eta'\leq1$, then $0\leq A_j(\eta')-A_j(\eta)\leq\eta'-\eta.$
Thus the Type I error is continuous and nondecreasing along the joined chain, with endpoint values zero and one. It therefore has a maximal admissible member whose Type I error is exactly $\varepsilon$. Denote any threshold representation by $(\widehat\tau,\widehat\eta)$. Let
\begin{equation*}
\kappa_{\mathcal C_0}:=\min_{\substack{P\in\mathcal C_0\\x\in\mathcal X}} P(x)>0,\qquad \kappa_{\mathcal C_1} :=\min_{\substack{Q\in\mathcal C_1\\x\in\mathcal X}} Q(x)>0.
\end{equation*}
If two induced test functions in the chain satisfy $\psi<\psi'$, then for some $x_0^n$ and $\delta>0$, $\psi'(x_0^n)-\psi(x_0^n)=\delta.$ Uniform full support gives
\begin{equation*}
\mathbb E_{P^{\otimes n}}[\psi']-\mathbb E_{P^{\otimes n}}[\psi]\geq\delta\kappa_{\mathcal C_0}^n >0 \forall P\in\mathcal C_0,\qquad \mathbb E_{Q^{\otimes n}}[1-\psi]-\mathbb E_{Q^{\otimes n}}[1-\psi'] \geq \delta\kappa_{\mathcal C_1}^n>0\forall Q\in\mathcal C_1.
\end{equation*}
Hence the Type I error increases strictly and the Type II error decreases strictly along distinct test functions in the chain. The maximal admissible member is therefore the unique restricted minimiser as a test function. Since it is admissible for the unrestricted problem, \eqref{eq:projected-threshold-type-two-value} follows.
\end{IEEEproof}

%###################################################
\subsection{Corollary~\ref{cor:simple-binary-tightening}}
\label[appendix]{app:simple-binary-threshold-tightening}

\begin{corollary}
\label{cor:simple-binary-tightening}
Let $\mathcal C_0=\{P\}$ and $\mathcal C_1=\{Q\}$, where $P$ and $Q$ have full support on a finite alphabet. Then, for every $\lambda\in(0,1)$, $h_\lambda^\star=\log\frac{Q}{P},$ independently of $\lambda$. Moreover, for every $n\geq1$ and $\varepsilon\in(0,1)$, the optimised test $\psi_{\widehat\tau,\widehat\eta}^\star$ from proposition~\ref{prop:projected-tightening} satisfies $\mathbb E_{Q^{\otimes n}}\left[1-\psi_{\widehat\tau,\widehat\eta}^\star\right]=\beta_n^\star(\varepsilon;P,Q).$
\end{corollary}

\begin{IEEEproof}
The only projected pair is $(Q_\lambda^\star,P_\lambda^\star)=(Q,P)$, so $h_\lambda^\star=\log Q/P$ and $S^\star=\log Q^{\otimes n} / P^{\otimes n}.$ The projected threshold family is therefore the family of randomised likelihood ratio tests. Its maximal member with Type I error $\varepsilon$ is the randomised NP test, which attains $\beta_n^\star(\varepsilon;P,Q)$.
\end{IEEEproof}

%###################################################
\subsection{\Cref{lem:weighted-tail-bounds}}
\label[appendix]{app:weighted-tail-bounds}

\begin{lemma}
\label{lem:weighted-tail-bounds}
Let $\mathcal X$ be finite, let $g:\mathcal X\to\mathbb R$ be nonconstant, and let $\mathfrak W$ be a compact class of full support probability distributions. For i.i.d. observations under $W\in\mathfrak W$, set
\begin{equation*}
G_n:=\sum_{i=1}^n g(X_i), \qquad m_W:=\mathbb E_W[g], \qquad \Delta_{n,W} :=G_n-nm_W.
\end{equation*}
For every $0<s_0<s_1<+\infty$, there exists $A<+\infty$ such that, for all $W\in\mathfrak W$, $s\in[s_0,s_1]$, $t\in\mathbb R$, and $n\geq1$,
\begin{align*}
\mathbb E_{W^{\otimes n}} \left[e^{-s(G_n-t)} \mathbbm 1\{G_n\geq t\}\right] \leq
\frac{A}{\sqrt n}, \qquad \mathbb E_{W^{\otimes n}}\left[e^{s(G_n-t)}
\mathbbm 1\{G_n<t\}\right] \leq \frac{A}{\sqrt n}.
\end{align*}
Moreover, for every $0\leq M_0<+\infty$, there exist $0<a_{M_0}<A_{M_0}<+\infty$ and $n_{M_0}<+\infty$ such that, for $n\geq n_{M_0}$ and $|u|\leq M_0\log n$,
\begin{align*}
a_{M_0}\frac{e^{-su}}{\sqrt n} \leq \mathbb E_{W^{\otimes n}} \left[e^{-s\Delta_{n,W}} \mathbbm 1\{\Delta_{n,W}\geq u\} \right]&\leq A_{M_0}\frac{e^{-su}}{\sqrt n},\\a_{M_0}\frac{e^{su}}{\sqrt n}
\leq \mathbb E_{W^{\otimes n}} \left[ e^{s\Delta_{n,W}} \mathbbm 1\{\Delta_{n,W}<u\} \right]&\leq A_{M_0}\frac{e^{su}}{\sqrt n},
\end{align*}
uniformly over $W\in\mathfrak W$ and $s\in[s_0,s_1]$. The same conclusions hold, after changing the constants, when strict and weak endpoints are interchanged.
\end{lemma}

\begin{IEEEproof}
Compactness, full support, and nonconstancy of $g$ give $0<v_0 \leq \operatorname{Var}_W(g) \leq v_1 < +\infty$ and a uniform bound on the centred third absolute moments. The Berry-Esseen inequality~\cite[Th.~44, p.~2329]{polyanskiy2010channel} therefore gives $B<+\infty$ such that
\begin{equation*}
\sup_{\substack{W\in\mathfrak W\\z\in\mathbb R}} \left| W^{\otimes n}\left(\frac{G_n-nm_W}{\sqrt{n\operatorname{Var}_W(g)}}\leq z \right)-\Phi(z) \right|\leq \frac{B}{\sqrt n}.
\end{equation*}
Consequently, for every fixed $L>0$ there is $A_L<+\infty$ such that
\begin{equation*}
\sup_{\substack{W\in\mathfrak W\\y\in\mathbb R}} W^{\otimes n}\left(y\leq G_n<y+L\right)\leq\frac{A_L}{\sqrt n}.
\end{equation*}
The same estimate holds for endpoint atoms because the Gaussian cdf is continuous and the Berry-Esseen bound also applies to strict cdfs. Partitioning $[t,+\infty)$ and $(-\infty,t)$ into intervals of length
$L$, and summing the resulting geometric series using $s\geq s_0$, proves the first two bounds. For the lower bounds, fix $M_0$ and write $\sigma_W^2 := \operatorname{Var}_W(g).$ Choose $H>0$ such that $\frac{H}{\sqrt{2\pi v_1}}>2B+2.$ For $|u|\leq M_0\log n$, the normalised endpoints $(u+H)/(\sigma_W\sqrt n)$ and $(u+2H)/(\sigma_W\sqrt n)$ converge to zero uniformly. Hence, for all sufficiently large $n$, the Gaussian probability of the corresponding interval is at least $(2B+1)/\sqrt n$. Berry-Esseen then gives
\begin{equation*}
W^{\otimes n} \left(u+H<\Delta_{n,W}\leq u+2H\right)\geq \frac{1}{\sqrt n}.
\end{equation*}
On this event, $e^{-s\Delta_{n,W}} \geq e^{-su-2s_1H},$ which proves the first lower bound. The interval
$u-2H<\Delta_{n,W}\leq u-H$ gives the second. The corresponding upper bounds follow from the first part with $t=nm_W+u$, after factoring out $e^{-su}$ or $e^{su}$. The atom estimate permits the stated changes of endpoints.
\end{IEEEproof}

%###################################################
\subsection{Proof of~\Cref{prop:polynomial-refinement}}
\label[appendix]{app:polynomial-refinement}

\begin{IEEEproof} 
Set $\lambda:=\lambda_r^\star$, $P^\star:=P_r^\star$, $Q^\star:=Q_r^\star$, and $h_r^\star:=\log(Q^\star/P^\star)$. Define
\begin{equation*} 
R(x):= \frac{Q^\star(x)^\lambda P^\star(x)^{1-\lambda}} {H_\lambda(Q^\star,P^\star)}.
\end{equation*} 
The proof of \Cref{thm:exact-composite-exponent} gives $D(R\|P^\star)=r$ and $D(R\|Q^\star)=\frac{1-\lambda}{\lambda} [D_\lambda(Q^\star\|P^\star)-r]$, and shows that $(Q^\star,P^\star)$ is projected at order $\lambda$. Writing $m:=\mathbb E_R[h_r^\star]$, we obtain 
\begin{align} m &= r-\frac{1-\lambda}{\lambda} \left[D_\lambda(Q^\star\|P^\star)-r\right], \label{eq:proof-polynomial-mean} \\ \log H_\lambda(Q^\star,P^\star) &= \lambda m-r = -(1-\lambda)m -\frac{1-\lambda}{\lambda} \left[D_\lambda(Q^\star\|P^\star)-r\right].
\label{eq:proof-polynomial-hellinger}
\end{align}
 
The proof of \Cref{thm:projected-finite-blocklength-achievability} therefore gives $\log\mathbb E_P[e^{\lambda h_r^\star}] \leq\log H_\lambda(Q^\star,P^\star)$ for every $P\in\mathcal C_0$, and $\log\mathbb E_Q[e^{-(1-\lambda)h_r^\star}] \leq\log H_\lambda(Q^\star,P^\star)$ for every $Q\in\mathcal C_1$. Define the tilted laws
\begin{equation*}
\begin{aligned} 
\widehat P(x) &:= P(x)\exp\left\{ \lambda h_r^\star(x) -\log\mathbb E_P[e^{\lambda h_r^\star}] \right\}, \\ \widehat Q(x) &:= Q(x)\exp\left\{ -(1-\lambda)h_r^\star(x) -\log\mathbb E_Q[e^{-(1-\lambda)h_r^\star}] \right\}.
\end{aligned}
\end{equation*} 
The two tilted classes are compact and have uniform full support. Moreover, $h_r^\star$ is nonconstant, since otherwise $P^\star=Q^\star$, contradicting $r<D(\mathcal C_1\|\mathcal C_0)$. Thus \Cref{lem:weighted-tail-bounds} applies uniformly. With $S_n:=\sum_{i=1}^n h_r^\star(X_i)$, change of measure and \eqref{eq:proof-polynomial-hellinger} give, for every $u\in\mathbb R$,
\begin{equation} 
P^{\otimes n}(S_n\geq nm+u) \leq C_{\mathcal C_0}n^{-1/2}e^{-nr-\lambda u} \qquad \forall P\in\mathcal C_0,
\label{eq:proof-polynomial-null-tail}
\end{equation}
and
\begin{equation}
\begin{aligned} 
Q^{\otimes n}(S_n<nm+u) \leq{}& C_{\mathcal C_1}n^{-1/2} \exp\left\{ -n\frac{1-\lambda}{\lambda} [D_\lambda(Q^\star\|P^\star)-r] +(1-\lambda)u \right\},
\end{aligned}
\label{eq:proof-polynomial-alternative-tail}
\end{equation}
uniformly over $Q\in\mathcal C_1$. Take $u_n:=-(2\lambda)^{-1}\log n+B$. By \eqref{eq:proof-polynomial-null-tail}, the test $\varphi_n^{(B)} :=\mathbbm 1\{S_n\geq nm-(2\lambda)^{-1}\log n+B\}$ has Type I error at most $C_{\mathcal C_0}e^{-\lambda B}e^{-nr}$. Choose $B$ so that $C_{\mathcal C_0}e^{-\lambda B}\leq1$. By \eqref{eq:proof-polynomial-mean}, its threshold is
\begin{equation*} 
n\left[ r-\frac{1-\lambda}{\lambda} (D_\lambda(Q^\star\|P^\star)-r) \right] -\frac{\log n}{2\lambda} +B.
\end{equation*}
Equation~\eqref{eq:proof-polynomial-alternative-tail} then gives
\begin{equation*}
\begin{aligned} 
\sup_{Q\in\mathcal C_1} \mathbb E_{Q^{\otimes n}}[1-\varphi_n^{(B)}] \leq{} C_{\mathcal C_1}e^{(1-\lambda)B} n^{-1/(2\lambda)} \exp\left\{ -n\frac{1-\lambda}{\lambda} [D_\lambda(Q^\star\|P^\star)-r] \right\}.
\end{aligned}
\end{equation*}
This proves the required upper order. For the converse, put $T_n:=S_n-nm$. The exact likelihood ratio identities are
\begin{equation*}
\begin{aligned} 
(P^\star)^{\otimes n}(A) &= e^{-nr} \mathbb E_{R^{\otimes n}} \left[e^{-\lambda T_n}\mathbbm 1_A\right], \\ (Q^\star)^{\otimes n}(A) &= \exp\left\{ -n\frac{1-\lambda}{\lambda} [D_\lambda(Q^\star\|P^\star)-r] \right\} \mathbb E_{R^{\otimes n}} \left[e^{(1-\lambda)T_n}\mathbbm 1_A\right].
\end{aligned}
\end{equation*}
Applied to $\{R\}$, \Cref{lem:weighted-tail-bounds} gives, for everyfixed $M<+\infty$, uniformly over $|u|\leq M\log n$,
\begin{equation*}
\begin{aligned} 
(P^\star)^{\otimes n}(T_n\geq u) &= \Theta\left( e^{-nr}n^{-1/2}e^{-\lambda u} \right), \\ (Q^\star)^{\otimes n}(T_n<u) &= \Theta\left( n^{-1/2} \exp\left\{ -n\frac{1-\lambda}{\lambda} [D_\lambda(Q^\star\|P^\star)-r] +(1-\lambda)u \right\} \right),
\end{aligned}
\end{equation*}
with the same orders for strict or weak endpoints. Consequently, there exist constants $B_-<B_+$ such that, for allsufficiently large $n$,
\begin{equation*}
\begin{aligned} 
(P^\star)^{\otimes n} \left( T_n\geq-\frac{\log n}{2\lambda}+B_- \right) &>e^{-nr}, \\ (P^\star)^{\otimes n} \left( T_n>-\frac{\log n}{2\lambda}+B_+ \right) &<e^{-nr}.
\end{aligned}
\end{equation*}
Hence the threshold of an exact randomised NP test lies between these two values, and
\begin{equation*} 
\beta_n^\star(e^{-nr};P^\star,Q^\star) = \Theta\left( n^{-1/(2\lambda)} \exp\left\{ -n\frac{1-\lambda}{\lambda} [D_\lambda(Q^\star\|P^\star)-r] \right\} \right).
\end{equation*} 
Pairwise reduction gives $\beta_n^\star(e^{-nr};\mathcal C_0,\mathcal C_1) \geq\beta_n^\star(e^{-nr};P^\star,Q^\star)$, which proves \eqref{eq:polynomial-refinement}. Taking logarithms gives
\begin{equation*}
\begin{aligned} -\frac{1}{n} \log\beta_n^\star(e^{-nr};\mathcal C_0,\mathcal C_1) ={}& \frac{1-\lambda_r^\star}{\lambda_r^\star} \left[ D_{\lambda_r^\star}(Q_r^\star\|P_r^\star)-r \right] + \frac{\log n}{2\lambda_r^\star n} + O\left(\frac{1}{n}\right).
\end{aligned}
\end{equation*}
\end{IEEEproof}

%##################################
\subsection{Proof of~\Cref{cor:polynomial-slope}}
\label[appendix]{app:polynomial-slope}

\begin{IEEEproof}
By \Cref{cor:hoeffding-tangent},
\begin{equation*}
\frac{d}{dr} \max_{0<\lambda<1}\frac{1-\lambda}{\lambda}\left[D_\lambda(\mathcal C_1\|\mathcal C_0)-r \right]=-\frac{1-\lambda_r^\star}{\lambda_r^\star}.
\end{equation*}
Therefore
\begin{equation*}
\frac{1}{2}\left[1-\frac{d}{dr}\max_{0<\lambda<1}\frac{1-\lambda}{\lambda}\left[D_\lambda(\mathcal C_1\|\mathcal C_0)-r \right]\right]= \frac{1}{2} \left[1+\frac{1-\lambda_r^\star}{\lambda_r^\star}\right]=\frac{1}{2\lambda_r^\star},
\end{equation*}
which proves the result.
\end{IEEEproof}

%###################################################
\section{Least favourability}

\subsection{Proof of \Cref{prop:projected-ordering}}
\label[appendix]{app:projected-ordering}

\begin{IEEEproof}
For each $n\geq1$, write $S^\star(x^n):=\sum_{i=1}^n h_\lambda^\star(x_i).$ The two ordering assumptions in \Cref{prop:projected-ordering}, together with preservation of stochastic order under independent sums, give
\begin{equation*} 
S^\star\text{ under }P^{\otimes n} \leq_{\rm st} S^\star\text{ under }(P_\lambda^\star)^{\otimes n}, \qquad S^\star\text{ under }Q^{\otimes n} \geq_{\rm st} S^\star\text{ under }(Q_\lambda^\star)^{\otimes n}.
\end{equation*}
The function $s\mapsto \mathbbm 1\{s>\tau\} + \eta\mathbbm 1\{s=\tau\}$ is nondecreasing, while its miss function is nonincreasing. Therefore
\begin{equation*} 
\mathbb E_{P^{\otimes n}}[\psi_{\tau,\eta}^\star] \leq \mathbb E_{(P_\lambda^\star)^{\otimes n}} [\psi_{\tau,\eta}^\star], \qquad \mathbb E_{Q^{\otimes n}}[1-\psi_{\tau,\eta}^\star] \leq \mathbb E_{(Q_\lambda^\star)^{\otimes n}} [1-\psi_{\tau,\eta}^\star].
\end{equation*} 
Since the projected distributions belong to their respective classes, taking the suprema gives the two equalities in proposition~\ref{prop:projected-ordering}.
\end{IEEEproof}

%###################################################
\subsection{Proof of \Cref{cor:projected-exact-reduction}}
\label[appendix]{app:projected-exact-reduction}

\begin{IEEEproof}
Every test satisfying the composite Type I constraint is admissible for
the selected simple pair. Hence
\begin{equation*} 
\beta_n^\star(\varepsilon;\mathcal C_0,\mathcal C_1) \geq \beta_n^\star(\varepsilon;P_\lambda^\star,Q_\lambda^\star).
\end{equation*} 
Conversely, the first equality in proposition~\ref{prop:projected-ordering} makes $\psi_{\tau,\eta}^\star$ admissible for the composite problem, while the second identifies its composite Type II error with its Type II error under $Q_\lambda^\star$. Its assumed optimality for the simple pair therefore gives the reverse inequality.
\end{IEEEproof}
%###################################################
\subsection{\Cref{lem:exponential-family-stochastic-order}}
\label[appendix]{app:exponential-family-stochastic-order}

\begin{lemma}
\label{lem:exponential-family-stochastic-order} Let $\{P_\theta:\theta\in\Theta\}$ be the natural exponential family introduced in \Cref{sec:exact-reduction}. For every $n\geq1$, $\theta_1<\theta_2$, and bounded measurable nondecreasing $f:\mathbb R\to\mathbb R$,
\begin{equation*} 
\mathbb E_{P_{\theta_1}^{\otimes n}}[f(T_n)] \leq \mathbb E_{P_{\theta_2}^{\otimes n}}[f(T_n)], \qquad T_n:=\sum_{i=1}^nT(X_i).
\end{equation*}
\end{lemma}

\begin{IEEEproof}
For $\theta_2>\theta_1$,
\begin{equation*} 
\frac{dP_{\theta_2}^{\otimes n}} {dP_{\theta_1}^{\otimes n}} = \exp\{(\theta_2-\theta_1)T_n -n[\psi(\theta_2)-\psi(\theta_1)]\} =:L(T_n),
\end{equation*} 
where $L$ is nondecreasing and has expectation one under $P_{\theta_1}^{\otimes n}$. If $U,U'$ are independent copies of $T_n$ under this law, then $2\operatorname{Cov}(f(U),L(U)) = \mathbb E[ (f(U)-f(U'))(L(U)-L(U')) ] \geq0.$ Since the difference of the two expectations in the lemma equals this covariance, the claim follows.
\end{IEEEproof}

%############################################
\subsection{Proof of \Cref{prop:endpoint-reduction}}
\label[appendix]{app:endpoint-reduction}

 \begin{IEEEproof} Suppose first that $\theta_+^{\mathcal C_0}<\theta_-^{\mathcal C_1}$. The likelihood ratio $dP_{\theta_-^{\mathcal C_1}}/dP_{\theta_+^{\mathcal C_0}}$ is increasing in $T$. By the NP lemma, an optimal endpoint test has the form 
$\varphi_n^{\rm e} = \mathbbm 1\{T_n>t\} + \eta\mathbbm 1\{T_n=t\},$ for $0\leq\eta\leq1. $  \Cref{lem:exponential-family-stochastic-order} and monotonicity of this test and its miss function give 
 \begin{equation*} 
 \sup_{P\in\mathcal C_0} \mathbb E_{P^{\otimes n}}[\varphi_n^{\rm e}] = \mathbb E_{(P_{\theta_+^{\mathcal C_0}})^{\otimes n}} [\varphi_n^{\rm e}] \leq \varepsilon, \qquad {\rm and} \qquad \sup_{Q\in\mathcal C_1} \mathbb E_{Q^{\otimes n}}[1-\varphi_n^{\rm e}] = \mathbb E_{(P_{\theta_-^{\mathcal C_1}})^{\otimes n}} [1-\varphi_n^{\rm e}]. 
 \end{equation*} 
 Thus the endpoint test is admissible for the composite problem and attains the simple endpoint error. Every test admissible for the composite problem is also admissible for the endpoint pair, giving the reverse inequality. If $\theta_+^{\mathcal C_1}<\theta_-^{\mathcal C_0}$, the same argument applied to $-T$ gives the lower-tail endpoint reduction.  It remains to identify the projected pair. For $0<\lambda<1$, 
 \begin{equation*} 
 \begin{aligned}
 D_\lambda \left( P_{\theta_{\mathcal C_1}} \middle\| P_{\theta_{\mathcal C_0}} \right) =\frac{1}{\lambda-1} \Big[ \psi( \lambda\theta_{\mathcal C_1} + (1-\lambda)\theta_{\mathcal C_0}) -\lambda\psi(\theta_{\mathcal C_1}) -(1-\lambda)\psi(\theta_{\mathcal C_0}) \Big].
 \end{aligned}
 \end{equation*} 
 Differentiation gives 
 \begin{equation*} 
 \frac{\partial D_\lambda}{\partial\theta_{\mathcal C_0}} = \psi'(\theta_{\mathcal C_0}) - \psi'( \lambda\theta_{\mathcal C_1} + (1-\lambda)\theta_{\mathcal C_0}),\qquad \frac{\partial D_\lambda}{\partial\theta_{\mathcal C_1}} = \frac{\lambda}{\lambda-1} \left[ \psi'( \lambda\theta_{\mathcal C_1} + (1-\lambda)\theta_{\mathcal C_0}) - \psi'(\theta_{\mathcal C_1}) \right]. 
 \end{equation*} 
 Since $\psi'$ is strictly increasing, when $\theta_{\mathcal C_0}<\theta_{\mathcal C_1}$ the divergence decreases with $\theta_{\mathcal C_0}$ and increases with $\theta_{\mathcal C_1}$. Thus the minimum occurs at $\theta_{\mathcal C_0}=\theta_+^{\mathcal C_0}$ and $\theta_{\mathcal C_1}=\theta_-^{\mathcal C_1}$, so the projected pair is $(P_{\theta_-^{\mathcal C_1}},P_{\theta_+^{\mathcal C_0}})$. When the order of the intervals is reversed, the derivative signs reverse and the projected pair is $(P_{\theta_+^{\mathcal C_1}},P_{\theta_-^{\mathcal C_0}})$. These are exactly the endpoint pairs stated in the proposition. \end{IEEEproof}

%##########################################
\section{Numerical evaluation of the bounds}
\label[appendix]{app:numerical-renyi}

This appendix gives details of the numerical optimisation used in \Cref{sec:numerical-illustrations}. 

For the affine classes $\mathcal C_0=\{P_s:s\in[0,1]\},$ and $\mathcal C_1=\{Q_t:t\in[0,1]\},$ the composite R\'enyi divergence is evaluated by the joint minimisation \begin{equation} D_\lambda(\mathcal C_1\|\mathcal C_0) = \min_{(s,t)\in[0,1]^2} D_\lambda(Q_t\|P_s).
\label{eq:appendix-composite-renyi}
\end{equation} 
Thus, for each fixed $\lambda$, the two class parameters are optimised jointly over $[0,1]^2$. We first evaluate the objective over the parameter square to locate candidate minima and then refine the best candidates by constrained numerical optimisation. The calculation is repeated from multiple starting points and checked on a finer parameter grid. The same procedure is used for $D_\lambda(\mathcal C_0\|\mathcal C_1)$ when the opposite divergence direction is required.
 
For the exponentially decreasing Type I constraint $\varepsilon_n=e^{-nr}$, \Cref{cor:threshold-rate} gives the critical rate
\begin{equation} 
r_{\rm c} = D(\mathcal C_1\|\mathcal C_0) = \min_{(s,t)\in[0,1]^2} D(Q_t\|P_s).
\label{eq:appendix-critical-rate}
\end{equation}
For the first affine ternary example, $r_{\rm c}=0.094,$ $s^\star=0,$ and $t^\star=0.639.$ The rates used in \Cref{fig:numerical-bounds} are $r=0.35r_{\rm c}=0.033$ and $r=1.5r_{\rm c}=0.142$.

For the final affine ternary example,
$D(\widetilde{\mathcal C}_1\| \widetilde{\mathcal C}_0) = 0.017,$ and $D(\widetilde{\mathcal C}_0\| \widetilde{\mathcal C}_1) = 0.016.$ 
For the achievable regime, we optimise the expression in \Cref{eq:exact-composite-exponent}. The optimisation over the R\'enyi order is performed outside the joint class optimisation. Thus, for every trial value $0<\lambda<1$, we first compute
\begin{equation*} 
(s_\lambda^\star,t_\lambda^\star) \in \operatorname*{arg\,min}_{(s,t)\in[0,1]^2} D_\lambda(Q_t\|P_s),
\end{equation*}
and then optimise the resulting scalar function of $\lambda$.

For the first affine ternary example with $r=0.35r_{\rm c}$, this gives $\lambda^\star=0.601,$ $s_{\lambda^\star}^\star=0,$ and $t_{\lambda^\star}^\star=0.602,$
with $D_{\lambda^\star}(\mathcal C_1\|\mathcal C_0)=0.056.$
 
For the converse regime, we optimise the expression in \Cref{eq:exact-composite-strong-converse-exponent}. Again, for every trial value $\lambda>1$, the class parameters are first optimised jointly over $[0,1]^2$. For the first affine ternary example with $r=1.5r_{\rm c}$, this gives
$\lambda^\star=1.224,$ $s_{\lambda^\star}^\star=0,$ and $t_{\lambda^\star}^\star=0.662,$
with $D_{\lambda^\star}(\mathcal C_1\|\mathcal C_0) = 0.116.$
 
For a general Type I constraint $\varepsilon_n$, the reverse finite sample converse is the bound in \Cref{thm:finite-blocklength-composite-renyi-converse}. For the numerical figures we also use the converse obtained from the opposite divergence direction,
\begin{equation} 
B_{n,\mathrm{fwd}} = \exp\left\{ \sup_{\lambda>1} \left[ \frac{\lambda}{\lambda-1}\log(1-\varepsilon_n) - nD_\lambda(\mathcal C_0\|\mathcal C_1) \right] \right\}.
\label{eq:appendix-forward-bound}
\end{equation} 
The displayed R\'enyi converse is the larger of this bound and the bound in \Cref{thm:finite-blocklength-composite-renyi-converse}. For each R\'enyi order, the corresponding composite divergence is obtained by joint optimisation over $(s,t)\in[0,1]^2$ before the outer optimisation over $\lambda$.

For the fixed and subexponential Type I constraints in \Cref{fig:fixed-subexponential-bounds}, we use the calibrated projected test of \Cref{prop:projected-tightening}. Since the projected pair depends on $\lambda$, for each candidate order $0<\lambda<1$ we first compute
\begin{equation*} 
(s_\lambda^\star,t_\lambda^\star) \in \operatorname*{arg\,min}_{(s,t)\in[0,1]^2} D_\lambda(\widetilde Q_t\|\widetilde P_s).
\end{equation*} 
The corresponding projected log likelihood ratio is then formed and its threshold and boundary randomisation are calibrated under the required Type I constraint. The resulting Type II error is evaluated over the complete alternative class, and the order giving the smallest value is selected. We use 27 candidate R\'enyi orders between $0.001$ and $0.99$, with increased resolution near zero and one. The converse orders are instead obtained by continuous one dimensional optimisation over $\lambda>1$.

All numerical calculations were implemented in Python using NumPy, SciPy, and Matplotlib. NumPy was used for array operations and polynomial representations, while SciPy was used for numerical optimisation, root finding, special functions, and the linear programmes through its HiGHS interface. The optimisation routines included constrained local optimisation, scalar minimisation, Brent root finding, and differential evolution for independent global checks. Matplotlib was used to generate the numerical figures, and independent blocklength calculations were parallelised using Python's \texttt{concurrent.futures} module.

%###################################################
\bibliographystyle{IEEEtran}
\bibliography{my}
%###################################################
\end{document}